\documentclass[preprint,apjl,dvipdfmx]{aastex}
\usepackage{natbib}
\shorttitle{Swing Amplification of Galactic Spirals}

\shortauthors{Michikoshi \& Kokubo}
\keywords{galaxies: kinematics and dynamics, galaxies:spiral, methods: analytical}

\begin{document}

\title{
  Swing Amplification of Galactic Spiral Arms: Phase Synchronization of
  Stellar Epicycle Motion
}
\author{
  Shugo Michikoshi\altaffilmark{1}, and Eiichiro Kokubo\altaffilmark{2}
}
\altaffiltext{1}{
  Center for Computational Sciences, University of Tsukuba, Tsukuba, Ibaraki 305-8577, Japan
}
\altaffiltext{2}{
  Division of Theoretical Astronomy, National Astronomical Observatory of
  Japan, Osawa, Mitaka, Tokyo 181-8588, Japan
}

\email{
  michikos@ccs.tsukuba.ac.jp, and kokubo@th.nao.ac.jp
}

\begin{abstract}
We revisit the swing amplification model of galactic spiral arms
 proposed by Toomre (1981).
We describe the derivation of the perturbation equation in detail and
 investigate the amplification process of stellar spirals.
We find that the elementary process of the swing amplification is the
 phase synchronization of the stellar epicycle motion.
Regardless of the initial epicycle phase, the epicycle phases of stars in
 a spiral are synchronized during the amplification.
Based on the phase synchronization, we explain the dependence of the
 pitch angle of spirals on the epicycle frequency.
We find the most amplified spiral mode and calculate its pitch angle,
wavelengths, and amplification factor, which are consistent with those
obtained by the more rigorous model based on the Boltzmann equation by
Julian and Toomre (1966).
\end{abstract}

\section{Introduction}
The origin and evolution of spiral arms in disk galaxies is a fundamental problem in astrophysics.
The classical theory on spiral arm dynamics is the Lin-Shu model \citep{Lin1964, Lin1966}. 
The Lin-Shu model postulates a quasi-stationary standing wave pattern that rotates around the galactic center with a constant pattern speed. 
In this model, spirals are long-lived and the so-called winding problem is avoidable.
However, the wave packet evolves with the group velocity, and it is finally absorbed at the Lindbrad resonances \citep{Toomre1969, Lynden-Bell1972}.
Thus, to maintain the density wave it requires some generating mechanisms such as WASER \citep{Mark1976} or $Q$-barrier \citep{Bertin1989, Bertin1989a}. 

With the recent progress of $N$-body simulations of spiral galaxies, the new picture of spiral arm formation and evolution was proposed.
Contrast to the Lin-Shu model, the spiral arms in $N$-body simulations are not stationary but transient and recurrent, appearing and disappearing continuously \citep{Sellwood1984, Baba2009, Sellwood2000, Sellwood2010, Fujii2011}. 
The basic process in this activity is so-called swing amplification \citep{Toomre1981}.
In a differentially rotating disk, a leading wave rotates to a trailing one because of the differential rotation. 
If Toomre's $Q$ is $Q=1 \mbox{--} 2$, the amplitude of the rotating wave is enhanced by the self-gravity \citep{Goldreich1965, Julian1966, Toomre1981}. 
With a perturber such as the corotating over-dense region, the stationary wave patterns are excited by the swing amplification \citep{Julian1966}. 
Even if there are no explicitly corotating perturbers, the small leading wave always exists since a disk consists of a finite number of stars.
Thus, without an explicit perturber, the trailing wave can grow spontaneously due to the swing amplification mechanism \citep{Toomre1991}.

We have been exploring the role of the swing amplification in the spiral arm formation and evolution in detail.
\cite{Michikoshi2014}  investigated the pitch angle of spiral arms using local $N$-body simulations.
They found that the pitch angle decreases with the shear rate.
This is consistent with the results of the global $N$-body simulations \citep{Grand2013, Baba2015}. 
Based on the linear theory of the swing amplification, they obtained the pitch angle and found it agrees with that obtained through $N$-body simulations.
\cite{Michikoshi2016} extended the previous study and investigated the radial and azimuthal wavelengths and the amplitude of spiral arms using $N$-body simulations.
They found that the dependencies of these quantities on the shear rate or the epicycle frequency agree well with those according to the linear theory of the swing amplification.
These quantitative results indicate that the swing amplification surely plays an important role in the spiral arm formation and evolution.

The $N$-body simulations that support the swing amplification mechanisms show the formation of the multi-arm spirals \citep{Fujii2011, Baba2013, Grand2013, Michikoshi2014}.
  Thus we should be careful to apply the swing amplification mechanism to the grand-design spirals.
  In addition, the swing amplification model is constructed based on the local approximation \citep{Julian1966, Toomre1981}. 
  Therefore, strictly speaking, the swing amplification mechanisms is not directly applicable to the global structure.
  Instead, we can apply it to the multi-arm spirals or flocculent spirals.
  However, it has been suggested that the short-term activities of the grand-design spirals in barred galaxies may be explained by the swing amplification \citep{Baba2015}.
  A further study is necessary to clarify the role of the swing amplification in the global structure.
  However, since the swing amplification itself is a general and fundamental mechanism in the various types of disks, understanding a physical process of the swing amplification is important.
  For example, the origin of the short-scale spiral arms in Saturn's ring, so-called self-gravity wakes, may be formed by the swing amplification \citep[e.g.,][]{Salo1995, Michikoshi2015}.

The physical process of the swing amplification is complicated because it relates with three fundamental elements, the self-gravity, the shear, and the epicycle oscillation.
To shed light on the physical process of the swing amplification, 
\cite{Toomre1981} introduced the simple model of the swing amplification.
This model is similar to the model proposed by \cite{Goldreich1965} except for the treatment of the velocity dispersion.
To introduce the effect of the velocity dispersion of stars, he used the reduction factor instead of the gas pressure term.
Hereafter we refer to this model as the GLBT model. 
He posited that the motion of a particle can be described by the simple oscillation equation with the variable frequency and argued that his model gives
the same result as that of the rigorous model based on the collisionless Boltzmann equation by \cite{Julian1966} (hereafter refereed to as JT model).
While the basic equation in the JT model is complicated, the GLBT model is simple and its dynamical behavior is easy to understand.
Using the GLBT model, the swing amplification has been explained in some review papers \citep{Athanassoula1984, Dobbs2014}.

At first glance, it seems that the GLBT model was derived using the equation of motion of a single particle in the rotational frame.
However, strictly speaking, it is impossible to derive the basic equation directly. 
As shown below, instead of the equation of motion of a single particle, we should begin with the Lagrange description of the hydrodynamic equation.
Furthermore, the original numerical calculation method was ambiguous.
The reduction factor was used for introducing the effect of the velocity dispersion, but no details of its treatment were given.
In some subsequent review papers, the amplification process was explained based on the GLBT model, but the derivation of the basic equation and the detailed numerical treatment were not described there \citep{Athanassoula1984, Dobbs2014}.
We show that the additional assumption on the vorticity perturbation is necessary for deriving the basic equation and the naive numerical treatment leads to breakdown of the model in the strong shear case such as a Keplerian rotation.

\cite{Toomre1981} compared the amplification factor by the GLBT model with that by the JT model 
and found that the dependence of the amplification factor on the azimuthal wavelength has a similar tendency for the flat rotation curve.
However, the comparison with the general shear rate was not performed.
\cite{Athanassoula1984} and \cite{Dobbs2014} performed the similar analyses with the general shear rate.
However, they did not compare them with the JT model.
Furthermore, in all previous works, the wavelength and the pitch angle for the maximum amplification were not investigated in detail.
Thus, we investigate the detailed dependence of the amplification factor on the epicycle frequency, the pitch angle, and wavelengths and compare them with those in the JT model.
This work is the first comprehensive comparison between JT and GLBT models.
In addition, we clarify the dynamical behavior of the solution in detail and find the synchronization of the epicycle phase that was not pointed out in the previous works explicitly.

The outline of this paper is as follows.
Section 2 examines the basic equation of the GLBT model.
In Section 3, we solve the equation numerically and investigate the most amplified wave.
In Section 4, we derive the pitch angle formula by the order-of-magnitude estimate.
Section 5 is devoted to a summary.

\section{Basic Equation \label{sec:var}}
We revisit the GLBT model of the swing amplification proposed by \cite{Toomre1981}.
In this model, the evolution of the spiral amplitude is described by the spring dynamics with the
variable spring rate.
However, he did not describe the complete derivation of the basic equation.
Thus we describe the derivation in detail. 

\subsection{Displacement and Gravitational Force}
We consider a small thin region of a galaxy and introduce a local rotating Cartesian coordinate $(x,y,z)$. 
The $x$-axis is directed radially outward, the $y$-axis is parallel to the direction of rotation, and the $z$-axis is normal to the $x$-$y$ plane.  

We investigate the evolution of a rotating wave in the Lagrangian description.
A particle located at $(x_\mathrm{i},y_\mathrm{i})$ at the initial time $t_\mathrm{i}$ moves to $(X_1(x_\mathrm{i},y_\mathrm{i},t),Y_1(x_\mathrm{i},y_\mathrm{i},t))$ at time $t$.
In the unperturbed state, the surface density of particles $\Sigma_0$ is uniform.
The unperturbed position of the particle at time $t$ is $(X_0(x_\mathrm{i},y_\mathrm{i},t),Y_0(x_\mathrm{i},y_\mathrm{i},t))$.
Since in the unperturbed state the self-gravity parallel to the $xy$ plane vanishes, the equation of motion is
\begin{eqnarray}
  0 &=& 2 \Omega \frac{\mathrm{D} Y_0}{\mathrm{D}t} + 4 \Omega A X_0, \nonumber  \\
  0 &=& - 2 \Omega \frac{\mathrm {D} X_0}{\mathrm{D}t}, \label{eq:eomap}
\end{eqnarray}
where $\Omega$ is the circular frequency, and $A$ is the Oort constant.
The derivation $ \mathrm{D}/ \mathrm{D}t$ is the Lagrangian derivative with respect to $t$, which means the time derivative with $x_\mathrm{i}$ and $y_\mathrm{i}$ fixed.
We assume that the unperturbed solution is a circular orbit and is described by
\begin{eqnarray}
  X_0(x_\mathrm{i},y_\mathrm{i},t) &=& x_\mathrm{i} \label{eq:unpqx}, \\
  Y_0(x_\mathrm{i},y_\mathrm{i},t) &=& y_\mathrm{i} - 2 A x_\mathrm{i} (t-t_\mathrm{i}). \label{eq:unpqy}
\end{eqnarray}

We consider the perturbation due to the displacement.
The perturbed displacement generally depends on the position and the time as $(\xi_x(x,y,t), \xi_y(x,y,t))$.
The particle located at $(x,y)$ in the unperturbed state moves to the position $(x+\xi_x(x,y,t),y+\xi_y(x,y,t))$, then the density changes due to the displacement.
Considering the mass conservation, the surface density $\Sigma(x,y,t)$ is described by the Jacobian determinant
\begin{equation}
  \Sigma(x + \xi_x(x,y,t),y + \xi_y(x,y,t)) =  \Sigma_0 \left(\frac{\partial (x+\xi_x,y+\xi_y)}{\partial (x,y)} \right) ^{-1} .
\end{equation}
If the displacement is sufficiently small, neglecting the higher order terms, we rewrite the density perturbation $\Sigma_1 = \Sigma - \Sigma_0$ as 
\begin{equation}
\Sigma_1(x,y,t) \simeq \Sigma_1(x+\xi_x,y+\xi_y,t) \simeq - \Sigma_0 \left( \frac{\partial \xi_x}{\partial x} + \frac{\partial \xi_y}{\partial y} \right).
\label{eq:dendiv}
\end{equation}

We consider a single plane wave.
At $t_\mathrm{i}$ we assume that the wave phase is $  k_{x\mathrm{i}} x_\mathrm{i} + k_{y\mathrm{i}} y_\mathrm{i}$
where $k_{x\mathrm{i}}$ and $k_{y\mathrm{i}}$ are the radial and azimuthal wavenumbers at $t_\mathrm{i}$.
Using $X_0$ and $Y_0$, we can rewrite the phase as
\begin{equation}
  k_{x\mathrm{i}} X_0 + k_{y\mathrm{i}} (Y_0+2 A X_0(t-t_\mathrm{i})) = (k_{x\mathrm{i}} + 2 A k_{y\mathrm{i}} (t-t_\mathrm{i})) X_0 + k_{y\mathrm{i}} Y_0, 
\end{equation}
which indicates that the wave number varies with time as
\begin{eqnarray}
  k_x(t) &=& k_{x\mathrm{i}} + 2 A k_{y \mathrm{i}}(t-t_\mathrm{i}) = 2 A k_{y \mathrm{i}} t, \\
  k_y    &=& k_{y\mathrm{i}} ,
\end{eqnarray}
where we define $t_\mathrm{i}$ by $k_x(0) = 0$.

\begin{figure}
  \begin{center}
  	\includegraphics[width = 0.5\textwidth] {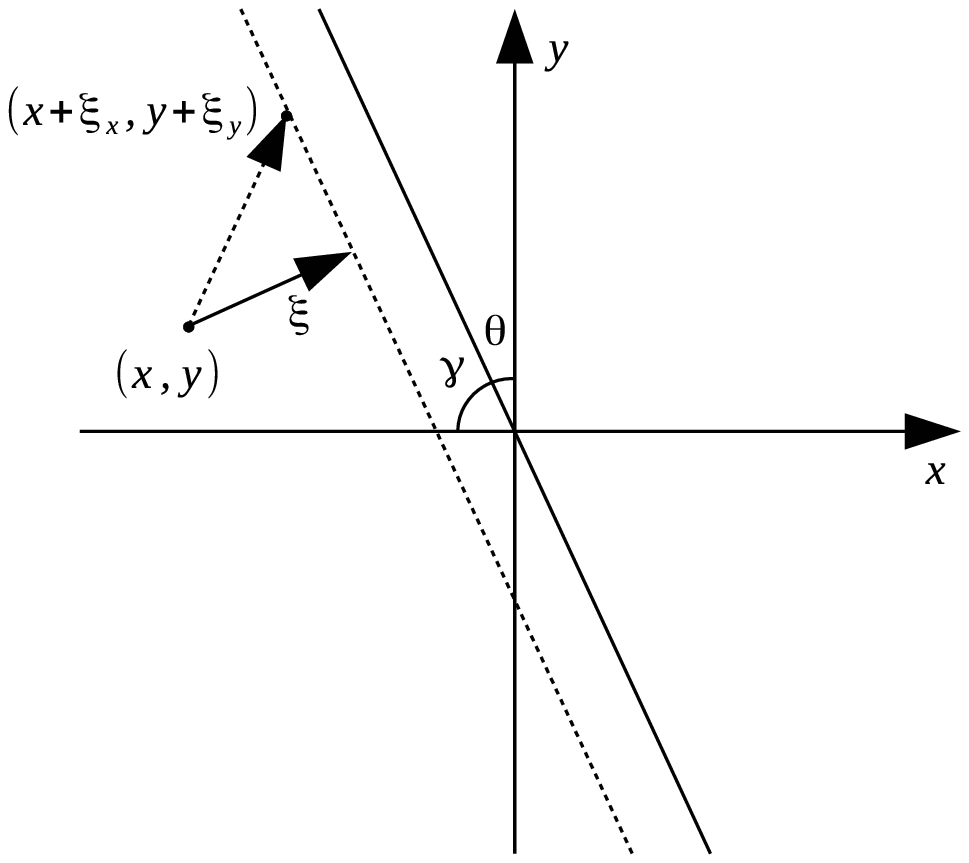}
  \end{center}
  \caption{Schematic illustration of a particle in a wave. The solid line denotes the density maximum line of the wave.
	The dotted line is the line parallel to the wave.
  The dotted arrow shows the displacement vector and the solid arrow shows the displacement normal to the wave $\xi$.
}
	  \label{fig:geofig}
\end{figure}

We assume the sinusoidal wave with the amplitudes $\xi_{x\mathrm{a}}$ and $\xi_{y\mathrm{a}}$
\begin{equation}
  \xi_x(x,y,t) = \xi_{x\mathrm{a}}(t) \exp(i(k_x(t) x + k_y y )),
  \label{eq:xix}
\end{equation}
\begin{equation}
  \xi_y(x,y,t) = \xi_{y\mathrm{a}}(t) \exp(i(k_x(t) x + k_y y )).
  \label{eq:xiy}
\end{equation}
The Poisson equation of the gravitational potential is
\begin{equation}
\nabla^2 \Phi = 4 \pi G \rho,
\label{eq:poi}
\end{equation}
where $\Phi$ is the gravitational potential, $\rho$ is the density.
The gravitational acceleration vector is 
\begin{equation}
(g(x,y,t) \sin \gamma(t), g(x,y,t) \cos \gamma(t)) = - \nabla \Phi,
\label{eq:gvec}
\end{equation}
where $g$ is the gravitational acceleration and $ \tan \gamma(t) = k_x(t)/k_y = 2At$.
The solution of Equation (\ref{eq:poi}) with the thin disk approximation is \citep{Toomre1969} 
\begin{eqnarray}
  \Phi_{1\mathrm{a}} = - \frac{2 \pi G \Sigma_{1\mathrm{a}}}{k},
  \label{eq:poisol}
\end{eqnarray}
where $\Phi_{1\mathrm{a}}$ and $\Sigma_{1\mathrm{a}}$ are the amplitudes of the gravitational potential perturbation and the surface density perturbation, respectively,
and $k(t)=\sqrt{k_x^2(t) + k_y^2}$.
Using Equations (\ref{eq:dendiv}), (\ref{eq:gvec}), and (\ref{eq:poisol}), we obtain \citep{Toomre1981}
\begin{equation}
  g(x,y,t) = 2 \pi G \Sigma_0 \xi(x,y,t)k(t),
  \label{eq:geq}
\end{equation}
where $\xi$ is the displacement normal to the wave (Figure \ref{fig:geofig})
\begin{equation}
  \xi(x,y,t) = \xi_x(x,y,t) \sin \gamma(t) + \xi_y(x,y,t) \cos \gamma(t).
  \label{eq:xieq_first}
\end{equation}
The negative and positive $\gamma$ corresponds to the leading and trailing waves, respectively.
If the wave is trailing, $\gamma$ relates to the pitch angle $\theta$ by $\gamma=90^\circ - \theta$.

Introducing the amplitude of the gravity $g_\mathrm{a}$ and the normal displacement $\xi_\mathrm{a}$, we obtain  
the amplitude relations from Equations (\ref{eq:geq}) and (\ref{eq:xieq_first}),
\begin{equation}
  g_\mathrm{a}(t) = 2 \pi G \Sigma_0 \xi_\mathrm{a}(t)k(t),
  \label{eq:gdef}
\end{equation}
\begin{equation}
  \xi_\mathrm{a} (t) = \xi_{x\mathrm{a}} (t) \sin \gamma(t) + \xi_{y\mathrm{a}}(t) \cos \gamma(t).
  \label{eq:xiadef}
\end{equation}

\subsection{Amplitude Equation}

Hereafter we omit the indication of the independent variables for each function if they are obvious. 

The displacement $(X_1(x_\mathrm{i},y_\mathrm{i},t),Y_1(x_\mathrm{i},y_\mathrm{i},t))$ obeys the equation of motion
\begin{eqnarray}
  \frac{\mathrm{D}^2 X_1}{\mathrm{D} t^2} &=&   2 \Omega \frac{\mathrm{D} Y_1}{\mathrm{D} t} + 4 \Omega A X_1 + g\sin \gamma, \label{eq:eomlagx} \\
  \frac{\mathrm{D}^2 Y_1}{\mathrm{D} t^2} &=& - 2 \Omega \frac{\mathrm{D} X_1}{\mathrm{D} t} + g \cos \gamma. \label{eq:eomlagy}
\end{eqnarray}
The following relation is always satisfied
\begin{equation}
  \frac{\mathrm{D}}{\mathrm{D} t} \exp(i(k_x(t) X_0(x_\mathrm{i},y_\mathrm{i},t) + k_y Y_0(x_\mathrm{i},y_\mathrm{i},t)) ) = 0.
\end{equation}
Thus, we have 
\begin{eqnarray}
  \frac{\mathrm{D}}{\mathrm{D} t} \xi_x(X_0(x_\mathrm{i},y_\mathrm{i},t), Y_0(x_\mathrm{i},y_\mathrm{i},t), t) &=& \frac{ \mathrm{d} \xi_{x\mathrm{a}}(t)}{\mathrm{d} t} \exp(i(k_x(t) X_0(x_\mathrm{i},y_\mathrm{i},t) + k_y Y_0(x_\mathrm{i},y_\mathrm{i},t)) ) ,\\
  \frac{\mathrm{D}^2}{\mathrm{D} t^2} \xi_x(X_0(x_\mathrm{i},y_\mathrm{i},t), Y_0(x_\mathrm{i},y_\mathrm{i},t), t) &=& \frac{ \mathrm{d}^2 \xi_{x\mathrm{a}}(t)}{\mathrm{d} t ^2} \exp(i(k_x(t) X_0(x_\mathrm{i},y_\mathrm{i},t) + k_y Y_0(x_\mathrm{i},y_\mathrm{i},t)) ).
\end{eqnarray}
We can obtain equations of the $y$-component in a similar way.

Using these relations and substituting $X_1(x_\mathrm{i},y_\mathrm{i},t)=X_0(x_\mathrm{i},y_\mathrm{i},t)+\xi_x(X_0(x_\mathrm{i},y_\mathrm{i},t),Y_0(x_\mathrm{i},y_\mathrm{i},t),t)$, $Y_1(x_\mathrm{i},y_\mathrm{i},t)=Y_0(x_\mathrm{i},y_\mathrm{i},t)+\xi_y(X_0(x_\mathrm{i},y_\mathrm{i},t),Y_0(x_\mathrm{i},y_\mathrm{i},t),t)$
into Equations (\ref{eq:eomlagx}) and (\ref{eq:eomlagy}), we obtain the amplitude equations 
\begin{eqnarray}
  \frac{\mathrm d^2 \xi_{x\mathrm{a}}}{\mathrm{d}t^2} &=& 2  \Omega \frac{\mathrm{d} \xi_{y\mathrm{a}}}{\mathrm{d}t} + 4 \Omega A \xi_{x\mathrm{a}} + g_\mathrm{a} \sin \gamma, \label{eq:pteomx} \\
  \frac{\mathrm d^2 \xi_{y\mathrm{a}}}{\mathrm{d}t^2} &=& -2 \Omega \frac{\mathrm{d} \xi_{x\mathrm{a}}}{\mathrm{d}t} + g_\mathrm{a} \cos \gamma.  \label{eq:pteomy} 
\end{eqnarray}

Eliminating $g_\mathrm{a}$ in Equations (\ref{eq:pteomx}) and (\ref{eq:pteomy}), we get
\begin{equation}
  \frac{\mathrm d^2 \xi_{x\mathrm{a}}}{\mathrm{d}t^2} - \tan \gamma \frac{\mathrm d^2 \xi_{y\mathrm{a}}}{\mathrm{d}t^2} = 2 \Omega \frac{\mathrm{d} \xi_{y\mathrm{a}}}{\mathrm{d}t} + 4 \Omega A \xi_{x\mathrm{a}} + 2 \Omega \tan \gamma \frac{\mathrm d \xi_{x\mathrm{a}}}{\mathrm{d}t}.
\end{equation}
Using $\mathrm{d} (\tan \gamma) / \mathrm{d} t = 2 A$, we rewrite the equation as
\begin{equation}
  \frac{\mathrm{d}}{\mathrm{d}t} \left(\frac{\mathrm d \xi_{x\mathrm{a}}}{\mathrm{d}t} - \tan \gamma \frac{\mathrm d \xi_{y\mathrm{a}}}{\mathrm{d}t} \right) = \frac{\mathrm{d}}{\mathrm{d}t} \left( 2 \Omega \tan \gamma \xi_{x\mathrm{a}} + 2 \Omega \xi_{y \mathrm{a}} - 2 A \xi_{y \mathrm{a}} \right). 
\end{equation}
Introducing the integral constant $C$, we obtain
\begin{equation}
  C = \frac{\mathrm d \xi_{x \mathrm{a}}}{\mathrm{d}t} - \tan \gamma \frac{\mathrm d \xi_{y\mathrm{a}}}{\mathrm{d}t}  -  2 \Omega \tan \gamma \xi_{x\mathrm{a}} + 2 B \xi_{y\mathrm{a}},
  \label{eq:cons}
\end{equation}
where $B=A-\Omega$ is the Oort constant.
This constant $C$ originates from the circulation theorems \citep{Goldreich1965}, which is proportional to the vorticity perturbation (See Appendix \ref{ap:vorticity} for details).

Using the variables $\xi_\mathrm{a}$ and $\xi_{x\mathrm{a}}$ and eliminating $\xi_{y\mathrm{a}}$ by Equation (\ref{eq:xiadef}), we rewrite Equations (\ref{eq:pteomx}) and (\ref{eq:pteomy}) as
\begin{eqnarray}
  \frac{\mathrm{d}^2 \xi_\mathrm{a}}{\mathrm{d}t^2} &=& g_\mathrm{a} -4 A ( - \Omega + \Omega \cos^2 \gamma + A \cos^4 \gamma) \xi_\mathrm{a} + 2\frac{ - \Omega + 2 A \cos^2 \gamma}{\cos\gamma} \left( \frac{\mathrm{d}\xi_{x\mathrm{a}}}{\mathrm{d}t} - \sin \gamma  \frac{\mathrm{d} \xi_\mathrm{a}}{\mathrm{d}t} \right) \label{eq:eomx2}, \\
  \frac{\mathrm{d}^2 \xi_{x\mathrm{a}}}{\mathrm{d}t^2} &=& g_\mathrm{a} \sin \gamma + 4 A \Omega \sin \gamma \xi_\mathrm{a} - \frac{2 \Omega}{\cos \gamma} \left(\sin \gamma \frac{\mathrm{d} \xi_{x\mathrm{a}}}{\mathrm{d}t} -  \frac{\mathrm{d} \xi_\mathrm{a}}{\mathrm{d}t} \right),
\end{eqnarray}
where we used $\mathrm{d} \gamma/ \mathrm{d} t = 2 A \cos^2 \gamma$.
Similarly, we rewrite Equation (\ref{eq:cons}) as
\begin{equation}
  \frac{\mathrm{d} \xi_{x\mathrm{a}}}{\mathrm{d} t} - \sin \gamma \frac{\mathrm{d} \xi_\mathrm{a}}{\mathrm{d}t} = C \cos^2 \gamma + 2 \cos \gamma (\Omega - A \cos^2 \gamma) \xi_\mathrm{a}.
\label{eq:cons2}
\end{equation}
Substituting Equations (\ref{eq:gdef}) and (\ref{eq:cons2}) into Equation (\ref{eq:eomx2}), we obtain
\begin{equation}
  \frac{\mathrm{d}^2 \xi_\mathrm{a}}{\mathrm{d}t^2} + \left(\kappa^2 - 8 A \Omega \cos^2 \gamma +12 A^2 \cos^4 \gamma - 2 \pi G \Sigma_0 k \right) \xi_\mathrm{a} = 
  2 C \cos \gamma (- \Omega + 2 A \cos^2 \gamma),
\label{eq:finalxi}
\end{equation}
where $\kappa$ is the epicycle frequency and we used $ 4A \Omega = 4\Omega^2 - \kappa^2$.
Equation (\ref{eq:finalxi}) with $C=0$ is the same as Equations (12) and (13) in \cite{Toomre1981}.

\begin{figure}
  \begin{center}
  	\includegraphics[width = 0.7\textwidth] {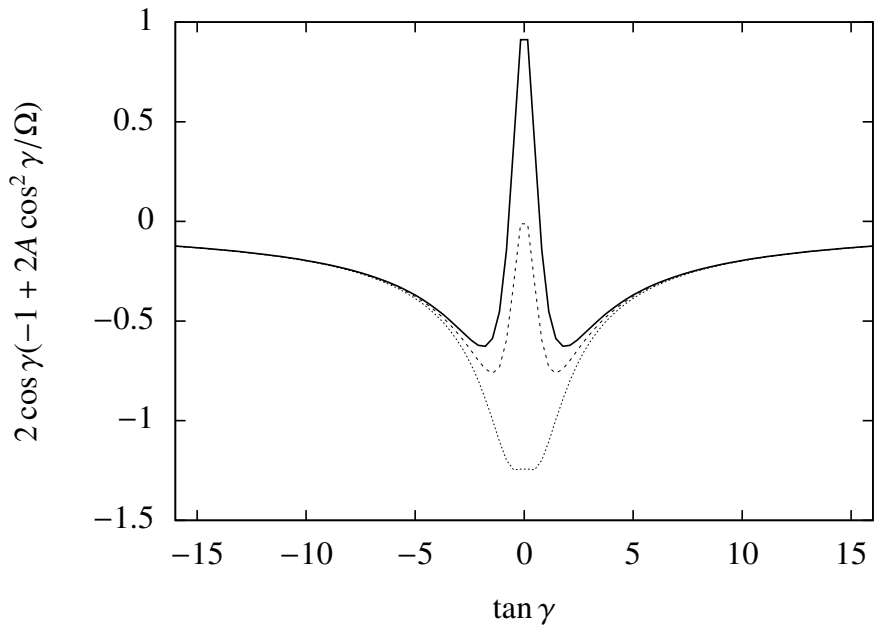}
  \end{center}
  \caption{Term in proportion to $C$ in Equation (\ref{eq:finalxi}) as a function of $\tan \gamma$ for $\kappa/\Omega=1.0$ (solid), $\kappa/\Omega=1.4$ (dashed), and $\kappa/\Omega=1.8$ (dotted).
	  \label{fig:vorticity}
}
\end{figure}

Figure \ref{fig:vorticity} shows the term in proportion to $C$ in Equation (\ref{eq:finalxi}).
When $\tan \gamma \sim 0$, the vorticity term can be large for small $\kappa$.
In all the previous works, $C=0$ was assumed explicitly \citep{Goldreich1965} or implicitly \citep{Toomre1981, Athanassoula1984, Dobbs2014}.
Following the previous works, we focus only on the case with $C=0$.
The condition $C=0$ is not always satisfied since it depends on the initial condition of the velocity perturbation.
The choice of $C=0$ means that we focus on the specified perturbation without the vorticity perturbation.

We summarize the amplitude equation without the effect of the velocity dispersion discussed so far.
The time evolution of $\xi_\mathrm{a}$ is described by the equation of the oscillation with the time variable frequency \citep{Goldreich1965, Toomre1981}
\begin{equation}
  \frac{\mathrm{d}^2 \xi_\mathrm{a}}{\mathrm{d}t^2} = - S \kappa^2  \xi_\mathrm{a},
\label{eq:xieq}
\end{equation}
where
\begin{equation}
  S = \frac{\kappa'^2 - 2 \pi G \Sigma_0 k}{\kappa^2},
  \label{eq:sp}
\end{equation}
\begin{equation}
\kappa'^2 =  \kappa^2 - 8 A \Omega \cos^2 \gamma +12 A^2 \cos^4 \gamma,
\label{eq:kappa0}
\end{equation}
where $S$ is the normalized squared frequency, which also can be interpreted as the spring rate of the system \citep{Toomre1981}.
While $k_y$ remains constant, $k_x$ evolves with time as $k_x = 2 A k_y t$.
Thus, $k=\sqrt{k_x^2 + k_y^2}$ also changes with time.
The frequency $\kappa'$ corresponds to the frequency of the oscillation without the self-gravity.

\subsection{Effect of Velocity Dispersion}
Equations (\ref{eq:xieq}) and (\ref{eq:sp}) do not include the effect of the stellar velocity dispersion, which reduces the effect of self-gravity.
The hydrodynamic model gives the similar equation \citep{Goldreich1965}:
\begin{eqnarray}
  S &=& \frac{\kappa'^2 - 2 \pi G \Sigma_0 k + c_\mathrm{s}^2 k^2}{\kappa^2},
\label{eq:gas}
\end{eqnarray}
where $c_\mathrm{s}$ is the sound velocity.  The term $c_s^2 k^2$ comes from the gas pressure. 

\cite{Toomre1981} introduced the effect of the stellar velocity dispersion by using the reduction factor $\mathcal{F}$
\begin{equation}
  S  = \frac{ \kappa'^2 - 2 \pi G \Sigma_0 k \mathcal{F}(s, \chi)}{\kappa^2},
\label{eq:spring0}
\end{equation}
\begin{equation}
  \mathcal{F} (s, \chi) = \frac{1-s^2}{\sin \pi s}\int_0^\pi \mathrm{d} \tau \exp(-\chi(1+\cos \tau)) \sin s \tau \sin \tau,
\end{equation}
where $\tau$ is the integral variable, $s$ and $\chi$ are
\begin{equation}
s =  \frac{\omega - m \Omega}{\kappa},
\end{equation}
\begin{equation}
\chi = \left(\frac{k \sigma_R}{\kappa} \right)^2,
\label{eq:chi_def}
\end{equation}
where $\omega$ is the wave frequency in the inertial frame, $m$ is the number of spiral arms, and $\sigma_R$ is the radial velocity dispersion \citep[e.g.,][]{Kalnajs1965, Lin1966, Lin1969, Binney2008}.
The properties of the reduction factor that are necessary in the following discussion are summarized in Appendix \ref{ap:red}.
The reduction factor means the degree of the reduction of the response of the disk to a perturbation due to the velocity dispersion. 
In terms of the physical meaning, the reduction factor should be $0 \leq \mathcal F \leq 1$. 

Strictly speaking, introducing $\mathcal{F}$ in this way was not justified by the Collisionless Boltzmann equation, which gives the integral equation \citep{Julian1966}.
The validity of using $\mathcal{F}$ is arguable because $\mathcal{F}$ was derived under the assumption of a tightly wound wave.
However, as shown by \cite{Toomre1981} this simple approach gives the similar results as those of the rigorous analysis by the collisionless Boltzmann equation.

\section{Results}

\subsection{Oscillation Frequency}

Using $\tilde \lambda_y = 2 \pi/(\lambda_\mathrm{cr} k_y)$ and $\lambda_\mathrm{cr}=4 \pi^2 G\Sigma_0/\kappa^2$, we rewrite Equation (\ref{eq:spring0}) as
\begin{equation}
  S = \frac{\kappa'^2}{\kappa^2} - \frac{\mathcal{F}(s, \chi)}{\tilde \lambda_y  \cos \gamma} ,
\label{eq:spring00}
\end{equation}
and Equation (\ref{eq:chi_def}) as
\begin{equation}
\chi = \left(\frac{3.36 Q}{2 \pi \tilde \lambda_y \cos \gamma} \right)^2,
\end{equation}
where $Q=\sigma_R \kappa/(3.36 G \Sigma_0)$ is Toomre's $Q$.
Note that $\tilde \lambda_y$ is $X$ in \cite{Toomre1981}.

To calculate $\mathcal{F}$, we need to specify $s$.
Since $s$ is the wave frequency in the rotating frame, the equation $s^2=S$ is satisfied.
Then, $S$ is given by the solution of the following equation
\begin{equation}
  S = \frac{\kappa'^2}{\kappa^2} - \frac{ \mathcal{F}(\pm \sqrt{S}, \chi)}{\tilde \lambda_y  \cos \gamma}.
\label{eq:spring}
\end{equation}

As shown in Appendix \ref{ap:exist}, for $\kappa/\Omega>2/\sqrt{3}$, Equation (\ref{eq:spring}) has two real solutions $S_{1}$ and $S_{2}$, 
which satisfy the inequality $S_{1} < \kappa'^2/\kappa^2 < S_{2}$.
From Equation (\ref{eq:sp}), the effect of the self-gravity reduces the frequency.
Thus, we take $S_1$.
We can numerically obtain $S_{1}$ by the bisection method or the relaxation method.
In general, the relaxation method is faster than the bisection method, but its convergence criterion is not trivial. 
Thus, in this paper, we adopt the bisection method, which assures the convergence.

For $1 \le \kappa/\Omega< 2/\sqrt{3}$ and small $\gamma \sim 0$, Equation (\ref{eq:spring}) does not have a real solution, 
where the negative reduction factor may arise from the breakdown of the tight-winding approximation.
Hence we introduce the lower bound to avoid the negative reduction factor and define the modified reduction factor $\mathcal{F}'$ as
\begin{equation}
  \mathcal{F}'(s, \chi) =
  \left\{
	\begin{array}{ll}
	  \mathcal{F}(s, \chi)  &   \mathcal{F}(s, \chi) > 0  \\
	  0  &   \mathcal{F}(s, \chi) \leq 0  \\
	\end{array}
	\right. .
\label{eq:spring2}
\end{equation}
Using this we define $S$ as
\begin{equation}
  S= \frac{\kappa'^2}{\kappa^2} - \frac{ \mathcal{F}'(\pm\sqrt{S}, \chi)}{\tilde \lambda_y \cos \gamma}.
\label{eq:springnew}
\end{equation}
Equation (\ref{eq:springnew}) always has a real solution that is less than or equal to $\kappa'^2/\kappa^2$ (see Appendix \ref{ap:exist}).

Figure \ref{fig:springg10q15} shows $S$ as a function of $\tan \gamma$, where $\kappa/\Omega=\sqrt{2}$ and $Q=1.5$.
We obtain $S$ by solving Equation (\ref{eq:springnew}) keeping the error less than $10^{-5}$.
If $|\tan \gamma|$ is not large, $\kappa'$ is less than $\kappa$.
The rotations of the local frame and the wave are in the opposite directions.
Thus, when the angular velocity of the wave rotation $|\mathrm{d} \gamma / \mathrm{d}t|$ is comparable to the circular velocity of the local frame $\Omega$, the comoving frame with the wave barely rotates against the inertial frame,
where the rotational effects such as the Coriolis force weakens.
Therefore, the oscillation frequency becomes small.
If $|\tan \gamma|$ is sufficiently large, $|\mathrm{d} \gamma / \mathrm{d}t|$ is negligible.
In this case, the comoving frame with the wave is approximately the same as the local rotational frame.
Then, without the self-gravity, we merely observe the usual epicycle motion with $\kappa' \simeq \kappa$.
If we consider the self-gravity, $S$ becomes small.  When $|\tan \gamma|$ is not large, $S$ is negative. Then, the wave is amplified.

\begin{figure}
	\plottwo{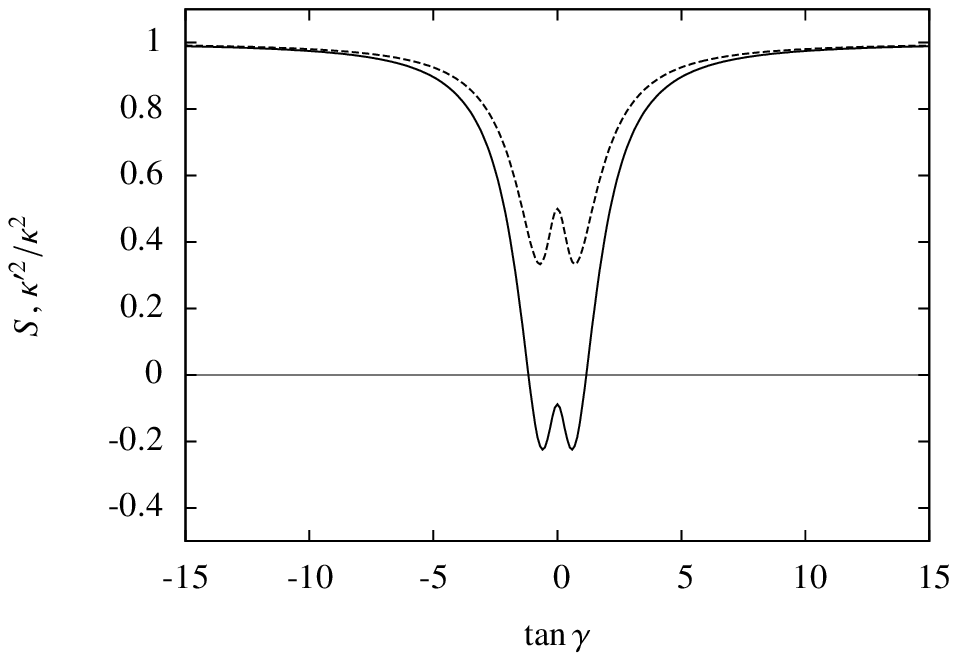}{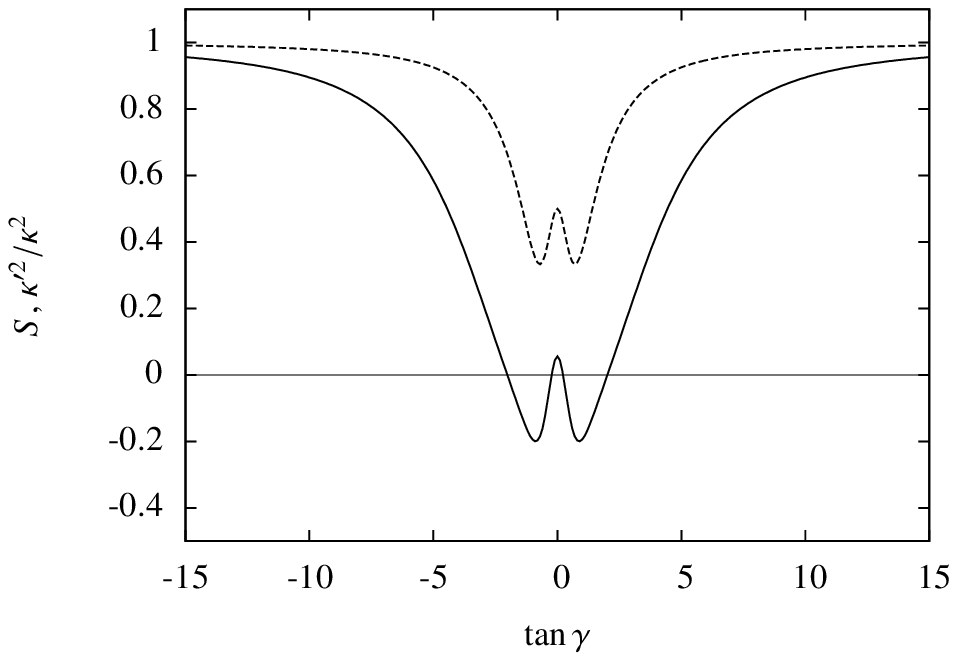}
	\caption{Squared frequencies $S$ (solid), and $\kappa'^2$ (dashed) as a function of $\tan \gamma$ with $\tilde \lambda_y=0.5$ (left panel) and $\tilde \lambda_y=2.0$ (right panel). The other parameters are $\kappa/\Omega=\sqrt{2}$ and $Q=1.5$.
	}
	\label{fig:springg10q15}
\end{figure}

\begin{figure}[htbp] 
\begin{tabular}{ccc}
    	\includegraphics[width=0.3\hsize] {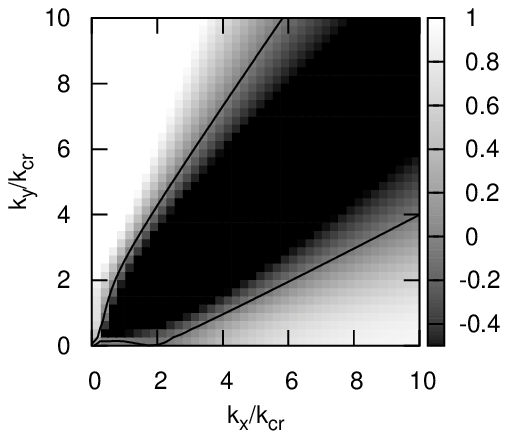} &
    	\includegraphics[width=0.3\hsize] {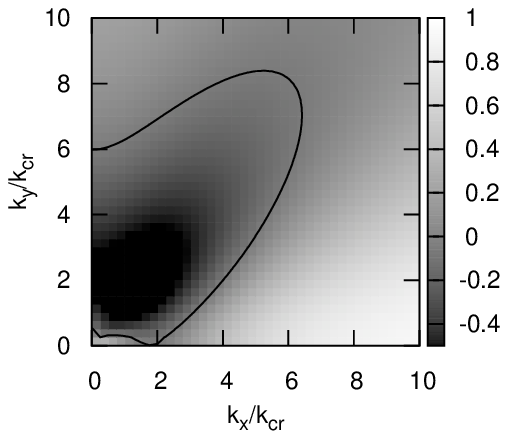} &
    	\includegraphics[width=0.3\hsize] {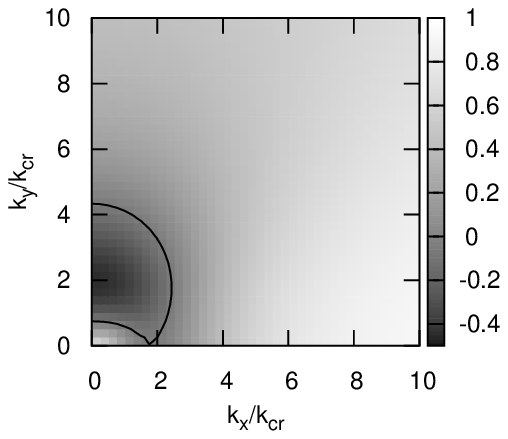} \\
    	\includegraphics[width=0.3\hsize] {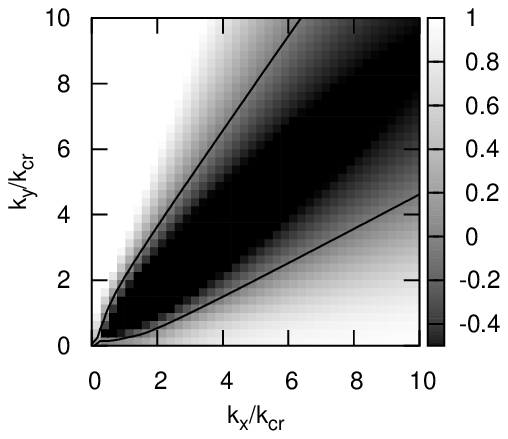} &
    	\includegraphics[width=0.3\hsize] {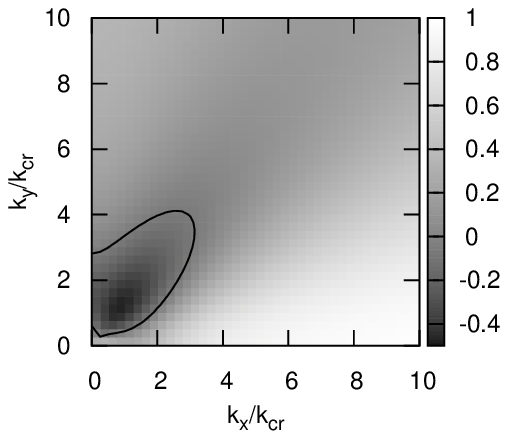} &
    	\includegraphics[width=0.3\hsize] {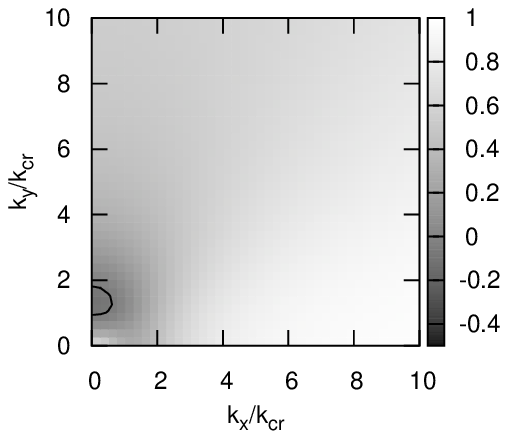} \\
    	\includegraphics[width=0.3\hsize] {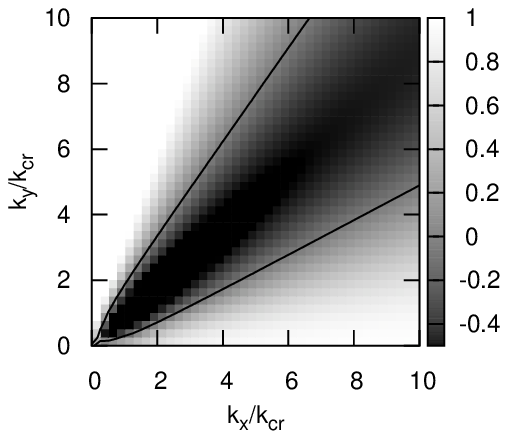} &
    	\includegraphics[width=0.3\hsize] {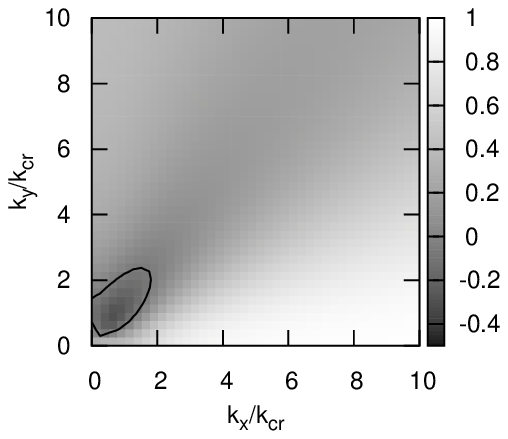} &
    	\includegraphics[width=0.3\hsize] {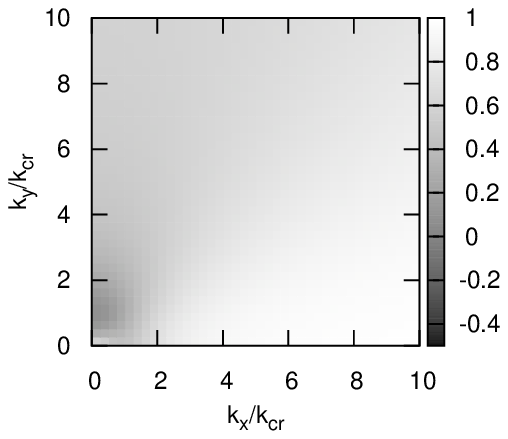} 
\end{tabular}
\caption{Squared frequency $S$ as a function of the wavenumbers $k_x$ and $k_y$. 
The epicycle frequencies are $\kappa/\Omega = 1.0$ (left), $1.4$ (middle), and $1.8$ (right), and the $Q$ values are $Q=1.0$ (top), $1.4$ (middle), and $1.8$ (bottom). 
The solid curve corresponds to $S=0$.
}
\label{fig:kxkydepa} 
\end{figure}

We evaluate $S$ as a function of $k_x$ and $k_y$ (Figure \ref{fig:kxkydepa}).
The wavenumbers $k_x$ and $k_y$ are related to $\tan \gamma$ and $\tilde \lambda_y$ as 
$k_x/k_\mathrm{cr} =\tan \gamma/\tilde \lambda_y$ and $k_y/k_\mathrm{cr} = 1/\tilde \lambda_y$ where $k_\mathrm{cr}= 2 \pi/\lambda_\mathrm{cr}$ is the critical wavenumber of the gravitational instability.
We can show from Equation (\ref{eq:springnew}) that $S$ is symmetric about the $k_y$ axes.
Thus we show the region where $k_x>0$ and $k_y>0$.
For $Q=1.0$ and $\kappa/\Omega=1.0$, the large area where $S<0$ exists.
In this parameter regime, the amplitude of the wave grows exponentially.
As $k_x$ and $k_y$ increase, $S$ approaches unity, that is, the motion for large wavenumbers is described by the epicycle oscillation.
For $\kappa/\Omega=1.4, 1.8$, if $k_y \simeq 0$ or $k_y$ is sufficiently large, the oscillation is stable with any $k_x$.
In order for the wave to be amplified extensively, it is necessary that $k_y$  should be a moderate value.
As $Q$ increases, the $S<0$ area shrinks.
This is because the self-gravity is suppressed by the velocity dispersion.
Similarly, as $\kappa/\Omega$ increases, the $S<0$ area shrinks.
The larger epicycle frequency may lead to weaker amplification in general.
This is consistent with the JT model for $Q \ge 1.4$.
In the JT model the amplification factor decreases with $\kappa/\Omega$ for $Q>1.4$ \citep{Michikoshi2016}.

\subsection{Time Evolution}

\begin{figure}
		\plotone{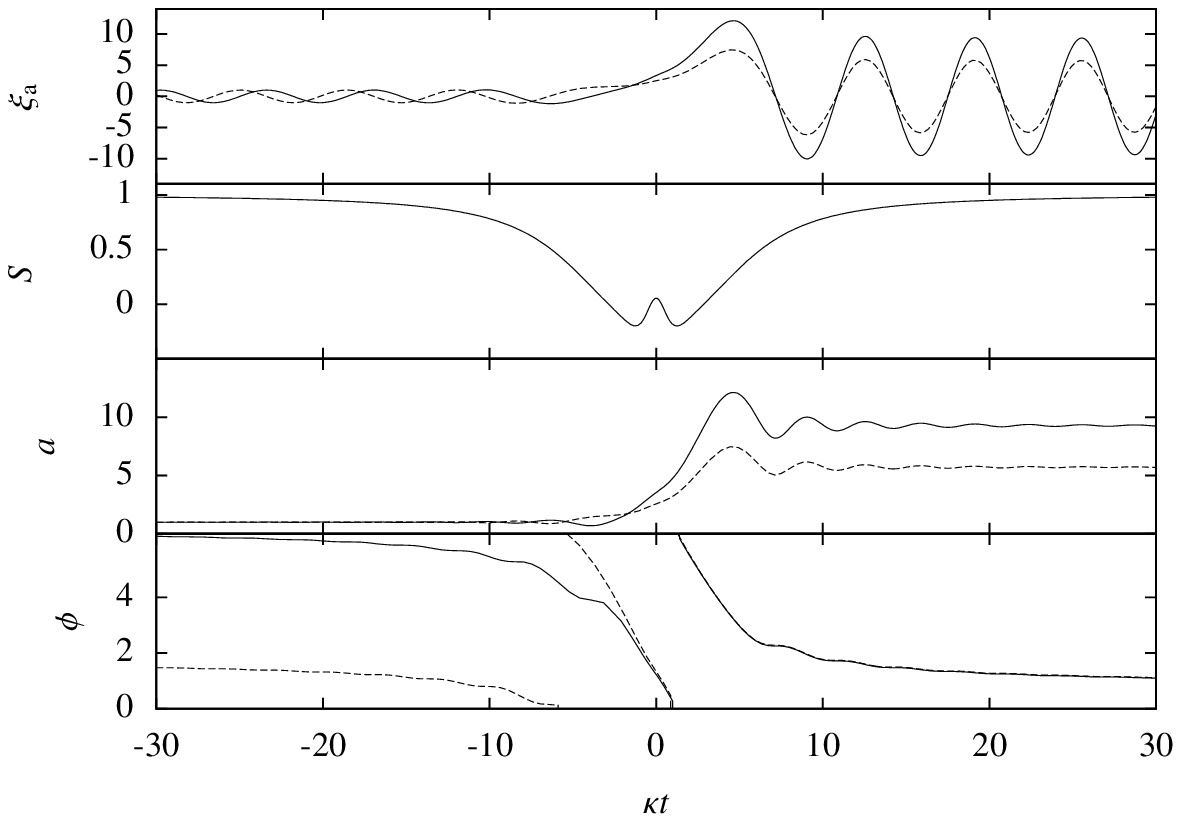}
		\caption{Time evolution of $\xi_\mathrm{a}$, $a$, and $\phi$ for $\kappa/\Omega=\sqrt{2}$ and $Q=1.5$, $\tilde \lambda_y=2.0$. The initial condition is $\xi_\mathrm{a} = 1$, $\frac{d \xi_\mathrm{a}}{dt}=0$ (solid) and $\xi_\mathrm{a} = 0$, $\frac{d \xi_\mathrm{a}}{dt}= -\kappa$ (dashed) at $t \kappa = -30$. }
\label{fig:xi_d_e}
\end{figure}

We solve Equation (\ref{eq:xieq}) with Equation (\ref{eq:springnew}) by the 4th-order Runge-Kutta method.
We define the osculating amplitude $a$ and phase $\phi$ as 
\begin{equation}
  a = \sqrt{\xi_\mathrm{a}^2 + \frac{1}{\kappa^2} \left(\frac{\mathrm{d} \xi_\mathrm{a}}{\mathrm{d} t} \right)^2},
\end{equation}
\begin{equation}
  \phi = - \kappa t - \tan^{-1} \left(\frac{1}{\xi_\mathrm{a} \kappa}\frac{\mathrm{d} \xi_\mathrm{a}}{\mathrm{d}t} \right) (\mathrm{mod}\, 2\pi),
\end{equation}
with which $\xi_\mathrm{a}$ is described as
\begin{equation}
\xi_\mathrm{a}(t) \simeq a \cos(\kappa t + \phi), 
\end{equation}
Figure \ref{fig:xi_d_e} shows the evolution of $\xi_\mathrm{a}$ for $\kappa/\Omega=\sqrt{2}$, $Q=1.5$, and $\tilde \lambda_y=2.0$.
For $\kappa t < -10$, $\sqrt{S}$ is approximated by $\kappa$. 
Thus, $\xi_\mathrm{a}$ approximately oscillates with period $2 \pi/\kappa$, and  $a$ and $\phi$ are almost constant.
For $-5 \lesssim \kappa t \lesssim 5$, $S$ decreases and becomes negative, where $a$ increases exponentially.
During the amplification, $\phi$ changes.
For $\kappa t > 10$, $\xi_\mathrm{a}$ approximately oscillates with period $2 \pi/\kappa$ again, and its amplitude is larger than that in the initial state.

We also calculate $\xi_\mathrm{a}$ with a different initial condition.
While the phase after the amplification is same, the final amplitude is different.
The amplification factor depends on the initial condition.

\subsection{Phase Synchronization and Pitch Angle}
We consider the dependence of the amplification factor on the initial condition.
We define $a_0$ and $\phi_0$ as the amplitude and phase at $\kappa t \to - \infty$, $a_1$ and $\phi_1$ as those for $\kappa t \to \infty$, and the amplification factor as $D_1 = a_1 /a_0$.
In general, the maximum value of $|\xi_\mathrm{a}|$ is larger than $a_1$.
The wave is most amplified to $|\xi_\mathrm{max}|$  at the first peak of the trailing wave. 
We define the maximum amplification factor as $D_2 = |\xi_\mathrm{max}| /a_0$.

\begin{figure}
  \plotone{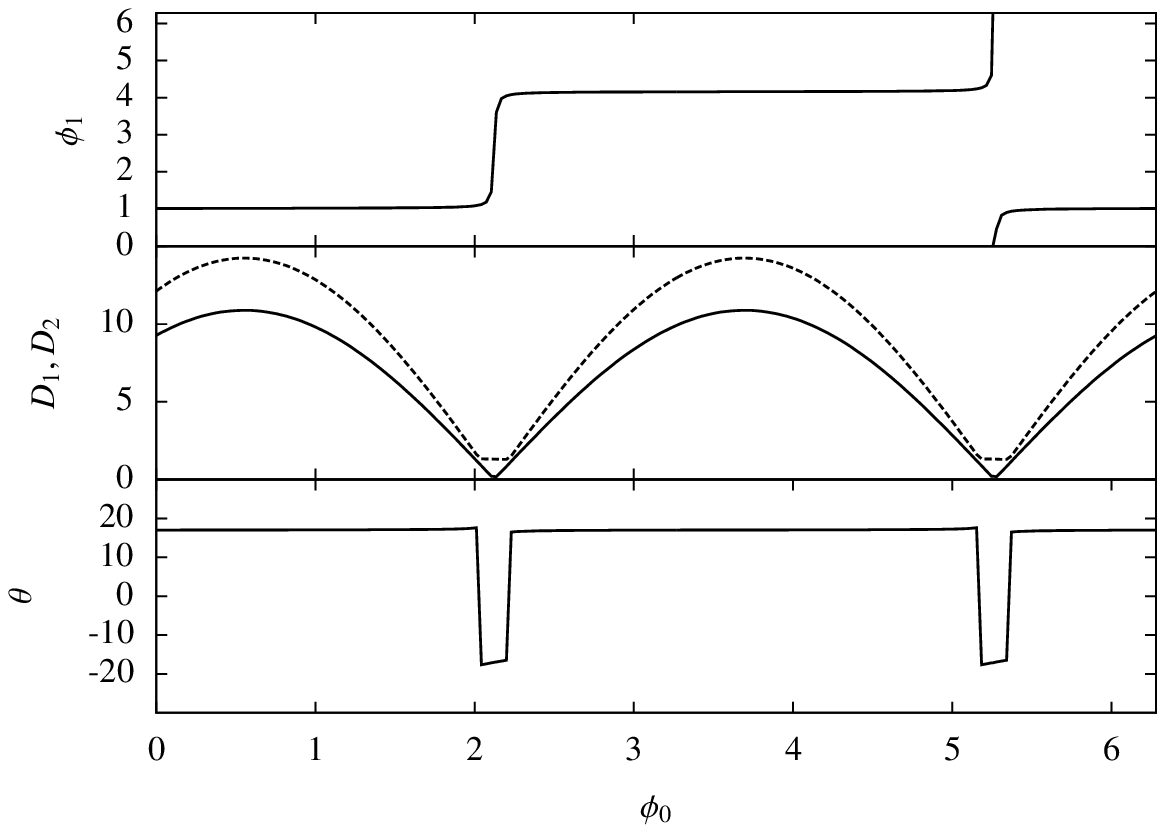}
  \caption{Resulting phase $\phi_1$ (top) and the amplification factors $D_1$ (solid) and $D_2$ (dashed) (middle) and the pitch angle (bottom) as a function of the initial phase $\phi_0$. The disk parameters are $\kappa / \Omega=1.4$,$Q=1.5$ and $\tilde \lambda_y=2.0$.}
\label{fig:phase_amp}
\end{figure}

Figure \ref{fig:phase_amp} shows the amplification factors $D_1$ and $D_2$, the resulting phase $\phi_1$ and the pitch angle $\theta$ of the trailing wave as a function of the initial phase $\phi_0$.
The resulting phase $\phi_1$ has approximately two values, $1.0$ and $4.2$, whose difference is about $\pi$, half an oscillation.
We find that the resulting phase is synchronized with the two discrete values independently of $\phi_0$.

We calculate the peak time $t_\mathrm{max}$ when $|\xi_\mathrm{a}|$ becomes the maximum, and calculate the pitch angle $\theta$ from
\begin{equation}
  \tan \theta = \frac{1}{2 A t_\mathrm{max}}.
\end{equation}
The bottom panel of Figure \ref{fig:phase_amp} shows the pitch angle.
The pitch angle does not depend on $\phi_0$, which is about $17.1^\circ$.
The synchronization of $\phi_1$ means that the peak time is independent of $\phi_0$ and thus the pitch angle does not depend on the initial condition.
Thus, the pitch angle is a function of only $\kappa$, $Q$, and $\tilde \lambda_y$.

\begin{figure}
		\plotone{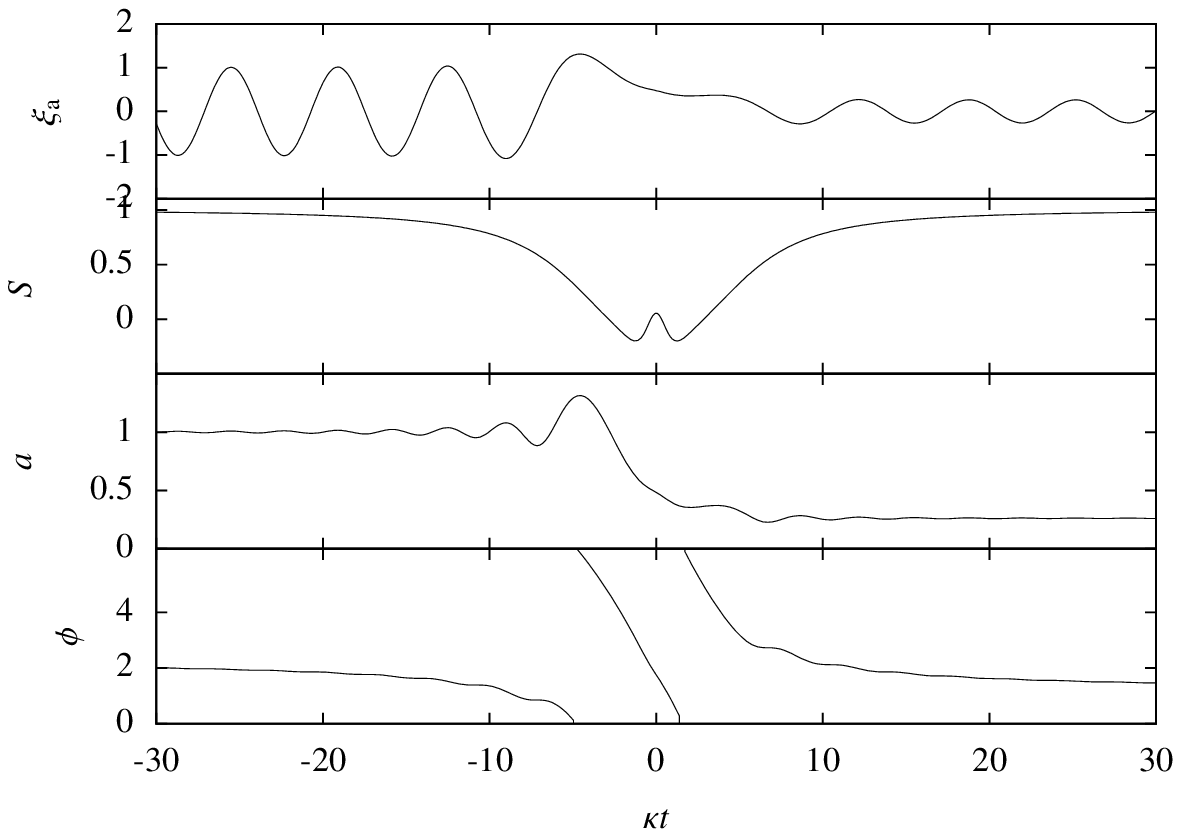}
		\caption{The same as Figure \ref{fig:xi_d_e} but for $\phi_0=2.10$.} 
		\label{fig:xi_d_ez}
\end{figure}

The pitch angle is negative in $2.04 < \phi_0 < 2.19$ and $5.18 < \phi_0 < 5.34$.
Figure \ref{fig:xi_d_ez} shows the time evolution of $\xi$ for $\phi_0=2.1$.
In this case, the amplification does not occur, and the final amplitude is smaller than the initial one.
Thus, $|\xi|$ has a maximum value at the negative time.
The corresponding pitch angle is negative.

The amplification factor depends on $\phi_0$. 
At $\phi_0 \simeq 0.56$ or $3.71$, the amplification factor has the maximum values $D_1=10.9$ and $D_2=14.2$.
On the other hand, at $\phi_0 \simeq 2.20, $ or $5.26$, the amplification factors are very small and the final amplitude is smaller than the initial one.

\newpage

\subsection{Most Amplified Wave}
\begin{figure}
	\plotone{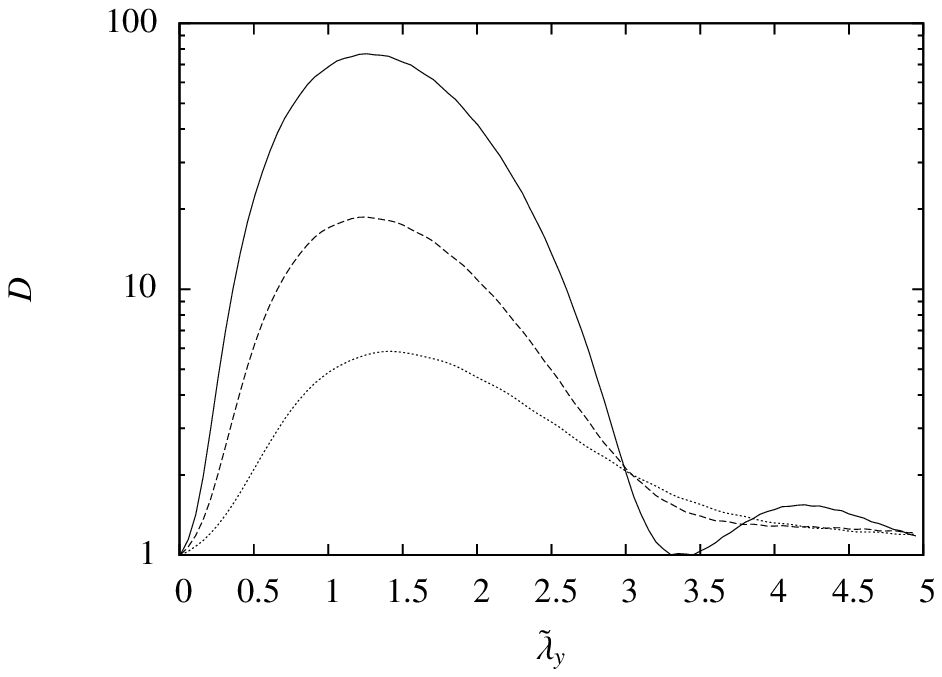}
	\caption{Amplification factor as a function of $\tilde \lambda_y$ for $Q=1.2$ (solid), $1.5$ (dash), and $2.0$ (dotted).
	  The shear rate is $\tilde \kappa=\sqrt{2}$.
	}
\label{fig:toomrefig7}
\end{figure}

\cite{Toomre1981} explored the dependence of the amplification factor on the normalized azimuthal wavelength $\tilde \lambda_y$.
We reproduce Fig. 7 in \cite{Toomre1981} as shown in Figure \ref{fig:toomrefig7}.
The same figure is also found in the review papers \citep{ Athanassoula1984, Dobbs2014}.
The amplification factor calculated here is similar but slightly larger than that in \cite{Toomre1981}. 
We cannot explain the difference completely since the detailed calculation method is not described in \cite{Toomre1981}.
In his calculation the initial phase might be $0$ or $\pi/2$, that is, the wave form for $t \kappa \to -\infty$ is $\sin \kappa t$ or $\cos \kappa t$.
In this paper, the initial phase is optimized to maximize the amplification factor. 
Thus, our amplification factor may be larger.

The amplification factor depends on $\tilde \lambda_y$.  The peak amplification factors are $D=81.2, 19.4,$ and $6.0$ for $\tilde \lambda_y=1.3, 1.3,$ and $1.4$, respectively.
Though the maximum amplification factor depends on $Q$ sensitively, the optimized $\tilde \lambda_y$ barely depends on $Q$.
The amplification factor becomes large for $1 \lesssim \tilde \lambda_y \lesssim 2$.
The optimized $\tilde \lambda_y$ also depends on $\kappa$ \citep{Athanassoula1984, Dobbs2014}.

We calculate the maximum amplification factor $D_\mathrm{max}$ by optimizing $\tilde \lambda_y$ for a disk with $\kappa$ and $Q$.
We define optimized $\tilde \lambda_y$ as $\tilde \lambda_{y,\mathrm{max}}$.
Then, we calculate the pitch angle $\theta_\mathrm{max}$ from the time when $\xi_\mathrm{a} = \xi_\mathrm{max}$, and evaluate the corresponding radial wavelength from 
\begin{equation}
  \tilde \lambda_{x,\mathrm{max}} = \frac{\lambda_{x,\mathrm{max}}}{\lambda_\mathrm{cr}} = \tilde \lambda_{y,\mathrm{max}} \tan \theta_\mathrm{max}.
\end{equation}

We compare these quantities with those in the JT model.
In the JT model, $\theta_\mathrm{JT}$, $\tilde \lambda_{x,\mathrm{JT}}$, $\tilde \lambda_{y,\mathrm{JT}}$, and $D_\mathrm{JT}$ are given as \citep{Michikoshi2016}
\begin{equation}
  \tan \theta_\mathrm{JT} = \frac{1}{2 \pi} \left(1+ \frac{2.095}{Q^{5.3}} \right)^{-1}\frac{\kappa}{A},
	\label{eq:fit_pitch2}
\end{equation}
\begin{equation}
	\tilde \lambda_{x,\mathrm{JT}}  = \frac{0.581Q^2-1.558Q+1.547}{1+2.095Q^{-5.3}} \frac{\Omega^2}{A \kappa},
	\label{eq:fit_xl} 
\end{equation}
\begin{equation}
  \tilde \lambda_{y,\mathrm{JT}}  = (3.653 Q^2 -9.789Q+9.721) \left(\frac{\Omega}{\kappa}\right)^2,
	    \label{eq:fit_xf2} 
\end{equation}
\begin{equation}
	D_\mathrm{max} = 0.0657 \exp\left( \frac{7.61}{Q} \right) \frac{\kappa A}{\Omega^2}.
	    \label{eq:fit_amp} 
\end{equation}

\begin{figure}
  \begin{center}
  	\includegraphics[width = 0.7\textwidth] {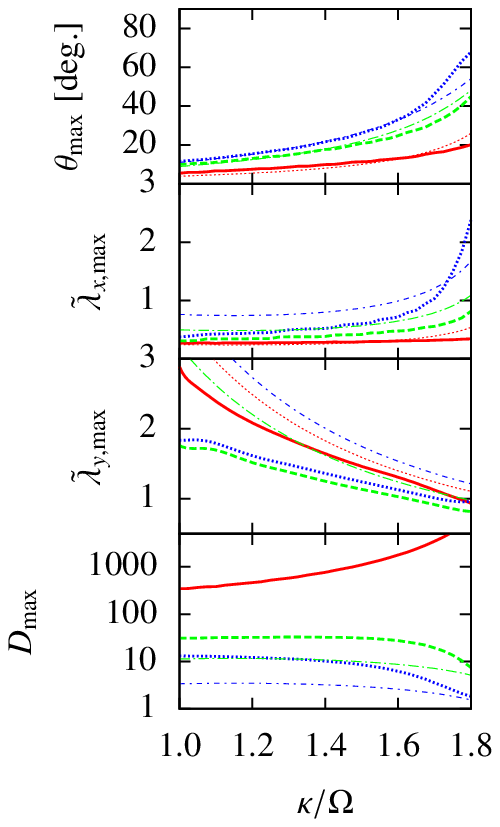}
  \end{center}
  \caption{Wave parameters $\theta_\mathrm{max}$, $\tilde \lambda_{x,\mathrm{max}}$, $\tilde \lambda_{y,\mathrm{max}}$ and $D_\mathrm{max}$ as a function of $\kappa / \Omega$ for $Q=1.0$ (red solid curve), $1.4$ (green dashed curve), and $1.8$ (blue dotted curve).   
	      The red dotted ($Q=1.0$), green dashed-dotted ($Q=1.4$), and blue short dashed-dotted  ($Q=1.8$) curves denote the fitting formulae in the JT model given by Equations (\ref{eq:fit_pitch2}), (\ref{eq:fit_xf2}), (\ref{eq:fit_xl}), and (\ref{eq:fit_amp}), respectively.
\label{fig:pitchall}
}
\end{figure}

Figure \ref{fig:pitchall} plots $\theta_\mathrm{max}$, $\tilde \lambda_{x,\mathrm{max}}$, $\tilde \lambda_{y,\mathrm{max}}$, and $D_\mathrm{max}$ against $\kappa$.
The overall tendency of the dependencies on $\kappa$ agrees with those by the JT model.
The pitch angle $\theta_\mathrm{max}$ increases with $\kappa$, which is consistent with the JT model.
The pitch angle $\theta_\mathrm{max}$ agrees well with $\theta_\mathrm{JT}$ for $\kappa/\Omega<1.6$, 
while for $\kappa/\Omega>1.6$, $\theta_\mathrm{max}$ is larger than $\theta_\mathrm{JT}$.
The dependence of $\tilde \lambda_{x,\mathrm{max}}$ on $\kappa/\Omega$ is similar to $\tilde \lambda_{x,\mathrm{JT}}$. 
While for $\kappa/\Omega \lesssim 1.6$, the dependence on $\kappa$ is weak, for $\kappa/\Omega \gtrsim 1.6$, $\tilde \lambda_{x,\mathrm{max}}$ increases with $\kappa/\Omega$.
However, $\tilde \lambda_{x,\mathrm{max}}$ is smaller than $\tilde \lambda_{x,\mathrm{JT}}$. 
The azimuthal wavelength $\tilde \lambda_{y,\mathrm{max}}$ decreases with $\kappa/\Omega$ and increases with $Q$.
Though the general trend is consistent with the JT model, $\tilde \lambda_{y,\mathrm{max}}$ is smaller than $\tilde \lambda_{y,\mathrm{JT}}$.
Especially, for $Q>1.4$ there is a large difference.
The behavior of $D_\mathrm{max}$ is in good agreement with $D_\mathrm{JT}$ (Equation (\ref{eq:fit_amp})), but its value is larger.

\section{Discussion}
The JT model is more rigorous than the GLBT model but is not easy to understand the wave dynamics from the evolution equation.
The GLBT model is less rigorous because of the intuitive introducing of the reduction factor that is valid for the tight winding approximation.
However it gives us an insight into a nature of the swing amplification.
These two models compliment one another.
In the GLBT model, we found that the final oscillation phase is independent of the initial oscillation phase.
This means that the oscillation phase is synchronized during the amplification. 
The essence of swing amplification is the phase synchronization of the epicycle motion.

\cite{Michikoshi2014} obtained the pitch angle by $N$-body simulation and linear analyses (JT model)
\begin{equation}
  \tan \theta = \frac{1}{7} \frac{\kappa}{A}.
\end{equation}
However, the physical interpretation has not yet been presented.
Based on the analyses in this paper, we describe the physical interpretation of the pitch angle formula.

We describe the swing amplification in terms of the phase synchronization of the epicycle motion.
We consider a single leading wave in a rotating frame.
The wave rotates from leading to trailing due to the shear with the angular speed $2\sin^2\theta A$ \citep[e.g.][]{Toomre1981, Binney2008}, while the rotating frame rotates in the opposite direction to the wave rotation with $\Omega$ in the inertial frame.
If the wave is tightly wound, that is, the pitch angle is small, the angular speed is very small.
Then the effects of the galactocentric rotation such as the Coriolis and tidal forces prevent the wave from the amplification due to the self-gravity.
When $\Omega \simeq 2 \sin^2 \theta A$, the rotations of the wave and the rotating frame are canceled out in the inertial frame.
 This happens when $\sin \theta = \sqrt{\Omega/2A}$, which roughly means that $\sin \theta$ is large such as $\theta \sim 90^\circ$.
 Then the stabilizing effects of the galactocentric rotation weaken and the self-gravity becomes relatively stronger \citep{Toomre1981, Dobbs2014}.
Therefore, particles are pulled toward the direction normal to the wave and their phases of the epicycle motion are synchronized, and consequently the wave density is amplified.

After the wave density reaches the maximum, it starts to decline.  At the same time, the
density of the other leading wave starts to grow, which reaches the maximum
quickly. This activity continues successively.  
Thus if we observe an entire disk, we
expect that the dominant wave corresponds to the wave with the maximum
amplification factor approximately.  
This hypothesis is supported by $N$-body simulations \citep{Michikoshi2014, Michikoshi2016}.  
They investigated the wave with the maximum amplification factor from the linear analysis and compared it with the dominant wave by $N$-body simulations.  
They confirmed that the time-averaged quantities of the dominant wave in $N$-body
simulations are consistent with those of the waves with the maximum amplification factor by
the linear analysis.
Therefore we assume that the observed spiral arms correspond to the wave with the maximum amplification factor.
The wave pitch angle $\theta$ decreases with time due to the shear as $\tan \theta = 1/ (2 A t)$ where $t$ is the time elapsed since $\theta=90^\circ$ \citep[e.g.,][]{Toomre1981, Binney2008}.
The observed pitch angle corresponds to the angle when the wave density reaches maximum after the epicycle phase is synchronized.
Roughly speaking, the synchronization starts when $\theta=90^\circ$.
The density maximization occurs on the timescale of the epicycle period $2 \pi/\kappa$.
Thus, substituting $t=2\pi/\kappa$ into $\tan \theta = 1/2At$, 
we obtain the pitch angle formula $\tan \theta \sim \kappa/A$ that is the same as Equation (\ref{eq:fit_pitch2}) except for the numerical factor and $Q$ dependence.

\section{Summary}
We considered the GLBT model introduced by \cite{Toomre1981}.
The formulation of the GLBT model is similar to that in \cite{Goldreich1965} except for the treatment of the gas pressure term.
We investigated the derivation and calculation procedure in detail.
To derive the basic equation, we need to assume that the constant of motion vanishes, which means that we focus on the initial perturbation without the vorticity perturbation \citep{Goldreich1965}.
We found that the GLBT model has the singularity and cannot be applied to the case where $\kappa/\Omega < 2/\sqrt{3}$.
To avoid this singularity, we introduced the lower bound in the reduction factor.

We calculated the maximum amplification factor and the corresponding wavelengths and compared them with those in the JT model.
The overall dependence on $\kappa/\Omega$ in the GLBT model is similar to that in the JT model.
However they are slightly different from those in the JT model.
In applying to the interpretation of numerical simulations or observations, we should use the GLBT model carefully.
It seems that the JT model is more reliable because it was derived in a more rigorous manner.
We have already confirmed that those in the JT model are in good agreement with those in $N$-body simulations  \citep{Michikoshi2016}.

Regardless of this drawback, the GLBT model is attractive because its basic equation is simple and gives us an insight into a nature of the swing amplification. 
Using the GLBT model, we found the synchronization phenomenon. 
The oscillation phase after the amplification is independent of the initial oscillation phase.
This is because the oscillation phase is synchronized during the amplification.
This may be the key process to understand the swing amplification.
Based on the phase synchronization, we derive the pitch angle formula by the order-of-magnitude discussion.
However, this process has not yet been confirmed by $N$-body simulations.
In the next paper, we will investigate the particle dynamics in spiral arms using $N$-body simulations.

We investigated the elementary process of the swing amplification.
However, in order to understand the overall spiral arm formation we should investigate the origin of the leading waves.
In the swing amplification, we postulate the existence of the strong leading waves.
If the leading waves come only from the particle noises, they are too small to account for the amplitudes of spiral arms.
Another mechanism is necessary to generate the strong leading waves.
One possible mechanism is the nonlinear wave-wave interaction \citep{Fuchs2005}.
However, the role of the nonlinear effect in the generation of the leading wave
is poorly understood.  In the future study, we will investigate this problem.

\appendix

\section{Vorticity Perturbation \label{ap:vorticity}}
We show the physical meaning of the constant $C$.
Using Equations (\ref{eq:unpqx}), (\ref{eq:unpqy}), (\ref{eq:xix}) and (\ref{eq:xiy}), we calculate the velocity field via the displacement vector
\begin{eqnarray}
  v_x(X_1,Y_1,t) &=& \frac{\mathrm{D} X_1}{\mathrm{D} t} = \frac{\mathrm{D} }{\mathrm{D} t} \left( X_0 + \xi_x(X_0,Y_0,t)\right) = \frac{\mathrm{d} \xi_{x\mathrm{a}}}{\mathrm{d}t} \exp(i(k_x X_0 + k_y Y_0)).\\
  v_y(X_1,Y_1,t) &=& \frac{\mathrm{D} Y_1}{\mathrm{D} t} = \frac{\mathrm{D} }{\mathrm{D} t} \left( Y_0 + \xi_y(X_0,Y_0,t)\right) = \frac{\mathrm{d} \xi_{y\mathrm{a}}}{\mathrm{d}t} \exp(i(k_x X_0 + k_y Y_0)) - 2 A X_0.
\end{eqnarray}
Introducing the variables $x=X_1=X_0+\xi_x$ and $y=Y_1=Y_0+\xi_y$ and neglecting the higher-order terms, we obtain
\begin{eqnarray}
  v_x(x,y,t) &\simeq& \frac{\mathrm{d} \xi_{x\mathrm{a}}}{\mathrm{d}t} \exp(i(k_x x + k_y y)), \\
  v_y(x,y,t) &\simeq& \frac{\mathrm{d} \xi_{y\mathrm{a}}}{\mathrm{d}t} \exp(i(k_x x + k_y y)) - 2 A (x-\xi_x) \nonumber \\
  &=&  \left( \frac{\mathrm{d} \xi_{y\mathrm{a}}}{\mathrm{d}t} + 2 A \xi_{x\mathrm{a}} \right) \exp(i(k_x x + k_y y)) - 2 A x. 
\end{eqnarray}
The $z$-component of the vorticity in the inertial frame is given as
\begin{eqnarray}
\omega_z = 2 \Omega +  \frac{\partial v_y}{\partial x} - \frac{\partial v_x}{\partial y} = 2 (\Omega- A) + \left(ik_x \left( \frac{\mathrm{d} \xi_{y\mathrm{a}}}{\mathrm{d}t} + 2 A \xi_{x\mathrm{a}} \right) - ik_y \frac{\mathrm{d} \xi_{x\mathrm{a}}}{\mathrm{d}t} \right)\exp(i(k_x x + k_y y)) ,
\end{eqnarray}
where $2 \Omega$ comes from the rotation of the coordinate system.
We divide the vorticity $\omega_z$ by the surface density $\Sigma=\Sigma_0+\Sigma_1$
\begin{equation}
  \frac{\omega_z}{\Sigma} \simeq \frac{2 (\Omega- A)}{\Sigma_0} -\frac{i k_y}{\Sigma_0} \left(
  \frac{\mathrm{d} \xi_{x\mathrm{a}}}{\mathrm{d}t}
  - \tan \gamma \frac{\mathrm{d} \xi_{y\mathrm{a}}}{\mathrm{d}t}
  - 2 \Omega \tan \gamma \xi_{x\mathrm{a}}
  +2 B \xi_{y\mathrm{a}} \right)\exp(i(k_x x + k_y y)) .
\end{equation}
Using Equation (\ref{eq:cons}), we obtain
\begin{equation}
  \frac{\omega_z}{\Sigma} \simeq \frac{2 (\Omega- A)}{\Sigma_0} -\frac{i k_y C}{\Sigma_0} \exp(i(k_x x + k_y y)) .
\end{equation}
Hence $C$ is in proportion to the amplitude of the perturbation of $\omega_z/\Sigma$.
Since there are no nonconservative forces in this two-dimensional system, the vorticity divided by the surface density moving with the wave remains constant with time.

\section{Reduction Factor \label{ap:red}}

We briefly summarize the mathematical properties of the reduction factor that are necessary for numerical calculation.
When $S=s^2>0$, the integral form of the reduction factor is \citep[e.g.,][]{Kalnajs1965, Lin1966, Lin1969, Binney2008}
\begin{equation}
  \mathcal{F} (\pm \sqrt{S}, \chi) = \frac{1-s^2}{\sin \pi s}\int_0^\pi \mathrm{d} \tau \exp(-\chi(1+\cos \tau)) \sin s \tau \sin \tau.
\label{eq:red}
\end{equation}
When $S<0$, $s$ is the pure imaginary number such as $s=iu$ where $u$ is the real number, and Equation (\ref{eq:red}) is written as
\begin{equation}
   \mathcal{F} (\pm i u, \chi) = \frac{1 + u^2}{\sinh \pi u }\int_0^\pi \mathrm{d} \tau \exp(-\chi(1+\cos \tau)) \sinh u \tau \sin \tau.
\label{eq:red2}
\end{equation}

\begin{figure}
  \plotone{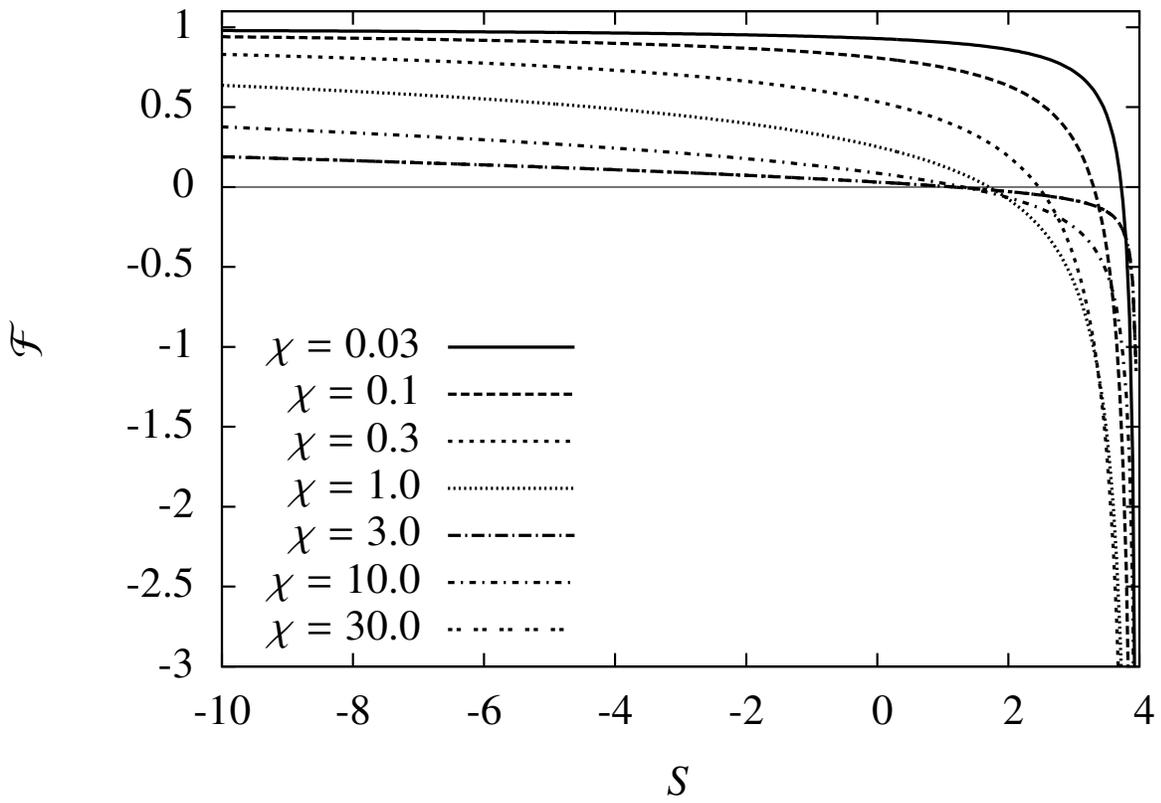}
	\caption{The reduction factor  $ \mathcal{F} $ as a function of $S$ for $\chi=0.03, 0.1, 0.3, 1.0, 3.0,$ and $10.0$. }
	\label{fig:redfuc}
\end{figure}

Figure \ref{fig:redfuc} shows $\mathcal{F}$ as a function of $S$.
The necessary properties of $\mathcal{F}$ that will be used in Appendix \ref{ap:exist} are the following:
  $ \mathcal{F} $ is continuous with respect to $S$ for $S<4$ (Appendix \ref{ap:red_domain}),
  $ \mathcal{F} $ is a monotonically decreasing concave function with respect to $S$ (Appendix \ref{ap:mono}),
  $ \mathcal{F} $ has the limit values $\lim_{S\to -\infty} \mathcal{F} = 1 $ and $\lim_{S\to 4-0} \mathcal{F} = -\infty $ (Appendix \ref{ap:asymp}), and
  $ \mathcal{F} $ is positive for $S<1$ with any $\chi$ (Appendix \ref{ap:red_pos}).
We give the proofs of these properties.

\subsection{Singularities \label{ap:red_domain}}
We consider the case where $S>0$.
Equation (\ref{eq:red}) has the singularities when $\sin \pi s = 0$, that is, $s$ is an integer.
We can show that Equation (\ref{eq:red}) with the limit $s \to 0$ and $s \to \pm 1$ is finite. 
Thus, they are the removable singularities at $s=0$ ($S=0$) and $s= \pm1$ ($S=1$).
For $s = \pm 2$ ($S=4$), the asymptotic form is 
\begin{equation}
  \mathcal{F} (2 - \epsilon, \chi) \simeq \frac{3 }{\pi \epsilon}\left(\int_0^\pi \mathrm{d} \tau \exp(-\chi(2+\cos \tau)) \sin 2 \tau \sin \tau + O(\epsilon) \right),
  \label{eq:limF}
\end{equation}
where $\epsilon$ is a sufficiently small value.
Since the integral is not equal to zero, the singularity cannot be removed. 
This is the essential singularity.
Therefore, $ \mathcal{F}$ diverges with the limit of $s\to \pm 2$ ($S\to 4$).
For $S<0$, there are no singularities because the denominator $\sinh \pi u$ is not equal to zero with $u\ne 0$.

The reduction factor $\mathcal{F}$ has the real finite value and is continuous for $S<4$.
If we allow the discontinuity and the singularities, we can define the reduction factor for the wider range of $S$.
However, the singular behavior seems to be unnatural.
Thus we consider only $S<4$.

\subsection{Monotonicity and Concavity  \label{ap:mono}}
The infinite series of the reduction factor is \citep{Binney2008}
\begin{equation}
  \mathcal{F} = \frac{2}{\chi } (1-S) e^{-\chi} \sum_{n=1}^\infty \frac{I_n(\chi)}{1-S/n^2},
  \label{eq:red_series1}
\end{equation}
and its first derivative with respect to $S$ is 
\begin{equation}
  \frac{\partial \mathcal{F}}{\partial S} = \frac{2 e^{-\chi}}{\chi} \sum_{n=1}^{\infty} \frac{I_n(\chi)(n^2-n^4)}{(n^2-S)^2},
\end{equation}
where $I_n$ is the modified Bessel function of the first kind.

Because all terms for $n\ge2$ in the infinite series are negative, the first derivative is negative.
The second derivative is
\begin{equation}
  \frac{\partial^2 \mathcal{F}}{\partial S^2} = \frac{2 e^{-\chi}}{\chi} \sum_{n=1}^{\infty} \frac{2n^2(1-n^2)I_n(\chi)}{(n^2-S)^3},
\end{equation}
where the term with $n=1$ is zero.  Since we consider $S<4$, we have $n^2-S > 0$ for $n \ge2$.
Thus all terms in the infinite series are negative, which means that the second derivative is also negative.
Therefore, for $S<4$, the inequalities $\partial \mathcal{F}/\partial S <0$ and $\partial^2 \mathcal{F}/ \partial S^2 <0$ are always satisfied.
This indicates that $\mathcal{F}$ is a monotonically decreasing concave function with $S$ for $S<4$.

\subsection{Asymptotic Behavior \label{ap:asymp}}
\subsubsection{$S\to4$}
We consider the behavior of $\mathcal{F}$ with the limit $S\to4$.
We rewrite the first term of Equation (\ref{eq:limF})
\begin{eqnarray}
  && \int_0^\pi \mathrm{d} \tau \exp(-\chi(2+\cos \tau)) \sin 2 \tau \sin \tau \nonumber \\ 
  &=& \int_0^{\pi/2} \mathrm{d} \tau \exp(-\chi(2+\cos \tau)) \sin 2 \tau \sin \tau +\int_{\pi/2}^{\pi} \mathrm{d} \tau \exp(-\chi(2+\cos \tau)) \sin 2 \tau \sin \tau \nonumber \\
  &=& \int_0^{\pi/2} \mathrm{d} \tau (\exp(-\chi(2+\cos \tau)) - \exp(-\chi(2-\cos \tau))) \sin 2 \tau \sin \tau < 0.
\end{eqnarray}
Thus, we obtain $ \lim_{S\to 4-0} \mathcal{F} = - \infty$ .

\subsubsection{$S\to  -\infty$}
Next we consider the behavior of $\mathcal{F}$ with the limit $S\to -\infty$.
From Equation (\ref{eq:red_series1}), with the sufficiently large $N_\mathrm{s}$, $\mathcal{F}$ is approximated by
\begin{equation}
  \mathcal{F} \simeq \frac{2}{\chi } (1-S) e^{-\chi} \sum_{n=1}^{N_\mathrm{s}} \frac{I_n(\chi)}{1-S/n^2}.
  \label{eq:inifnite}
\end{equation}
If $|S| \gg N_\mathrm{s}$, Equation (\label{eq:inifnite}) is approximated as
\begin{equation}
  \mathcal{F} \simeq \frac{2}{\chi } e^{-\chi} \sum_{n=1}^{N_\mathrm{s}} n^2 I_n(\chi).
  \label{eq:red_nega_inf}
\end{equation}

For arbitrary $\theta$, the following relation exists \citep{Binney2008}
\begin{equation}
  \exp(\chi \cos \theta) = \sum_{n=-\infty}^{\infty} I_n(\chi) \cos n \theta.
\end{equation}
Differentiating twice with respect to $\theta$ and substituting $\theta=0$, we obtain
\begin{eqnarray}
  \sum_{n=-\infty}^{\infty} n^2 I_n(\chi) &=& \chi e^\chi, \\
\end{eqnarray}
Thus, for $N_\mathrm{s} \to \infty$ and $S \to -\infty$ Equation (\ref{eq:red_nega_inf}) becomes
\begin{equation}
  \mathcal{F} \simeq \frac{2}{\chi } e^{-\chi} \times \frac{\chi e^\chi}{2} = 1.
\end{equation}
The reduction factor converges to unity as $S \to -\infty$.

Similarly, we can prove the first derivative $\partial \mathcal{F}/\partial S$ converges to zero as $S \to -\infty$.

\subsection{Sufficient Condition for Positive Reduction Factor \label{ap:red_pos}}
As shown in Figure \ref{fig:redfuc}, it seems that $\mathcal{F}$ is always positive for $S<1$.
From Equation (\ref{eq:red2}), we obtain 
\begin{equation}
 \mathcal{F}(\pm 1,\chi) = \lim_{S\to 1} \mathcal{F} = \frac{2}{\pi} \int_0^\pi \mathrm{d} \tau \exp(-\chi(1+\cos \tau)) \sin^2 \tau.
\label{eq:red11}
\end{equation}
Since the integrand in Equation (\ref{eq:red11}) is positive, $\mathcal{F}$ with $S=1$ is positive.
Considering that $\mathcal{F}$ is a monotonically decreasing function of $S$, we find that $\mathcal{F}$ is positive for $S<1$.

\section{Solution to Equation (\ref{eq:spring}) \label{ap:exist}}
In Appendix \ref{ap:ex1}, we show that Equation (\ref{eq:spring}) has two real solutions, where one of them is smaller than $\kappa'/\kappa$ if $\kappa/\Omega>2/\sqrt{3}$.
If $\kappa/\Omega<2/\sqrt{3}$ and $|\tan \gamma|$ is small, Equation (\ref{eq:spring}) has no real solutions.
This is caused by the negative reduction factor. 
Physically $\mathcal{F}$ should be $0\le \mathcal{F} \le 1$.
In Appendix \ref{ap:ex2} we show that Equation (\ref{eq:spring}) has a real solution regardless of $\kappa$ if we introduce the lower bound of the reduction factor to avoid the negative value.  

\subsection{Sufficient Condition for Existence of Solution \label{ap:ex1}}

We define $\delta$ as
\begin{equation}
  \delta(S) = \frac{\kappa'^2}{\kappa^2}- \frac{\mathcal{F}( \sqrt{S}, \chi)}{\tilde \lambda_y \cos \gamma} - S.
\label{eq:spring2}
\end{equation}
The function $\delta$ is continuous for $S<4$ because $\mathcal{F}$ is continuous for $S<4$ (Appendix \ref{ap:red_domain}).
Equation $\delta=0$ is the same as Equation (\ref{eq:spring}).
As shown in Appendix \ref{ap:asymp}, since the limit values of the reduction factor are  $ \lim_{S \to -\infty} \mathcal{F} = 1$ and $ \lim_{S \to 4} \mathcal{F} = -\infty$, we obtain the limit values of $\delta$ as $\lim_{S \to -\infty} \delta = \infty$ and $\lim_{S \to 4-0} \delta = \infty$. 
Thus, if $S$ where $\delta(S) <0$ exists, $\delta(S)=0$ has multiple real solutions because of the continuity.   

The first and second derivatives are
\begin{equation}
  \frac{\partial \delta}{\partial S} = - \frac{1}{\tilde \lambda_y \cos \gamma} \frac{\partial \mathcal{F}}{\partial S}- 1,
\end{equation}
\begin{equation}
  \frac{\partial \delta^2}{\partial S^2} = - \frac{1}{\tilde \lambda_y \cos \gamma} \frac{\partial^2 \mathcal{F}}{\partial S^2}.
\end{equation}
For $S\to4-0$, we have $\lim_{S\to4-0} \partial \delta/\partial S = \infty$.
On the other hand, for $S\to -\infty$, we have $ \lim_{S\to4-0} \partial \delta/\partial S = -1$ (Appendix \ref{ap:asymp}).
Thus, the real solution of $\partial \delta /\partial S =0$ exists.
Due to $\partial^2 \mathcal{F}/\partial S^2<0$ for $S<4$ (Appendix \ref{ap:mono}), the second derivative is positive $\partial^2 \delta/\partial S^2>0$, that is,
$\partial \delta /\partial S$ is a monotonically increasing function of $S$.
From the monotonicity of $\partial \delta /\partial S$, the equation $ \partial \delta /\partial S =0$ has the one real solution, that is, $\delta(S)$ is the only one local minimum.
Therefore, if $S$ where $\delta <0$ exists, Equation (\ref{eq:spring2}) has two real solutions.   

Substituting $S=\kappa'^2/\kappa^2$ into Equation (\ref{eq:spring2}), we obtain
\begin{equation}
  \delta(\kappa'^2/\kappa^2) = - \frac{\mathcal{F}( \kappa'/\kappa, \chi)}{\tilde \lambda_y \cos \gamma}.
\end{equation}
If $\mathcal{F}$ with $S=\kappa'^2/\kappa^2$ is positive, that is $\delta(\kappa'^2/\kappa^2)<0$, $\delta(S)$ has two real solutions: $S_1, S_2$, which satisfy the inequality $S_1 < \kappa'^2/\kappa^2 < S_2$.

As shown in Appendix \ref{ap:red_pos} when $S<1$, $\mathcal{F}$ is always positive.
Thus, the sufficient condition for the existence of the solution is $\kappa'^2/\kappa^2<1$.
From Equation (\ref{eq:kappa0}), $\kappa'^2/\kappa^2$ is always less than unity if  $\kappa/\Omega>2/\sqrt{3}$.
Hence, the sufficient condition for the existence of the solution is $\kappa /\Omega > 2/\sqrt{3}$.

\begin{figure}
	\plotone{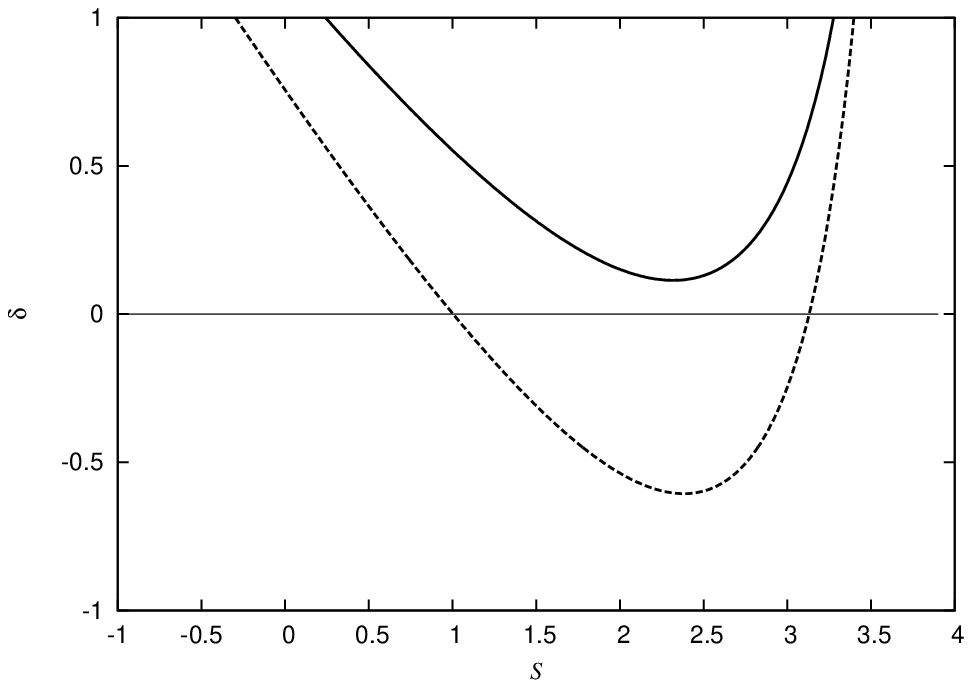}
	\caption{The value $\delta$ as a function of $S$ for $\kappa/\Omega=1$ ,$Q=1$, and $\tan \gamma=0$. The solid and dashed curves correspond to $\tilde \lambda_y=0.2$ and $\tilde \lambda_y=0.5$, respectively. The points at the intersection with $\delta=0$ give the solution of Equation (\ref{eq:spring}).}
	\label{fig:springnudif}
\end{figure}

If $\kappa/\Omega<2/\sqrt{3}$ and $|\tan \gamma|$ is small, $\kappa'^2/\kappa^2$ can be larger than unity.
For example, for $\kappa/\Omega=1$, $\kappa'^2/\kappa^2$ is larger than unity at $\tan \gamma \simeq 0$.
Then, the equation $\delta(S)=0$ may have no real solutions.
Figure \ref{fig:springnudif} shows the dependence of $\delta$ on $S$.
The parameters are $\kappa/\Omega=1$, $Q=1$ and $\tan \gamma=0$. For $\tilde \lambda_y=0.5$, $\delta(S)=0$ has the two real solutions, $S_1=1.00$ and $S_2=3.13$.
On the other hand,  $\delta(S)=0$ has no real solutions for $\tilde \lambda_y=0.3$.

\subsection{Modified Reduction Factor \label{ap:ex2}}
If the reduction factor is positive, the solution always exists.
The negative reduction factor might be unnatural, which means that the sign of the self-gravity term changes and the resulting frequency is larger than $\kappa'$.
Although the velocity dispersion is considered, it would be natural that the gravity reduces the frequency.
To avoid the negative reduction factor, we introduce the lower bound
\begin{equation}
  \mathcal{F}' =
  \left\{
	\begin{array}{ll}
	  \mathcal{F}  &  ( \mathcal{F} > 0)  \\
	  0  &  ( \mathcal{F} \leq 0)  \\
	\end{array}
	\right. .
\label{eq:spring3}
\end{equation}
We redefine $\delta(S)$ using $\mathcal{F}'$.
From the definition, $\delta(\kappa'^2/\kappa^2)$ is not positive.
On the other hand, for $S\to -\infty$, because of $\mathcal{F}' \to 1$, we have $\delta(S)\to \infty$.
Thus, $\delta(S)=0$ must have a solution that is less than or equal to $\kappa'^2/\kappa^2$.
When the original equation has a real solution, the modified equation has the same solution as that of the original equation.
When the original equation does not have a real solution, the modified equation has a solution that is equal to $\kappa'^2/\kappa^2$.

\end{document}